\documentclass[11pt]{amsart}

\usepackage{amssymb}
\usepackage{amsfonts}
\usepackage{amsmath}
\usepackage{amsxtra}
\usepackage{datetime}
\usepackage{latexsym}
\usepackage{caption}
\usepackage{cite}
\usepackage[usenames, dvipsnames]{color} 
\usepackage{csquotes}
\usepackage{dsfont}
\usepackage{empheq}
\usepackage{etoolbox}
\usepackage{enumerate}
\usepackage{float}
\usepackage{listings}
\usepackage{mathtools}
\usepackage{mathrsfs}
\usepackage{mleftright}
\usepackage{pgfplots}
\usepackage{stackengine}
\usepackage[caption=false]{subfig}
\usepackage{thmtools,thm-restate}
\usepackage{tikz}
\usepackage{url}
\usepackage{ytableau}
\usepackage[]{graphics}
\usepackage{bbm}
\usetikzlibrary{arrows,automata, pgfplots.fillbetween, matrix,positioning,fadings,calc,shapes,backgrounds,trees, plotmarks, patterns}
\newtheorem*{definition}{Definition}

\newtheorem{theorem}{Theorem}[section]
\newtheorem{corollary}{Corollary}[theorem]

\DeclareMathOperator{\RE}{Re}

\DeclareMathOperator{\mom}{MoM}

\DeclareMathOperator{\lca}{lca}
\DeclareMathOperator{\lcl}{lcl}

\newcommand{\cN}{\mathcal{N}}

\DeclarePairedDelimiter\abs{\lvert}{\rvert}%
\DeclarePairedDelimiter\norm{\lVert}{\rVert}%

\makeatletter
\let\oldabs\abs
\def\abs{\@ifstar{\oldabs}{\oldabs*}}
\let\oldnorm\norm
\def\norm{\@ifstar{\oldnorm}{\oldnorm*}}
\makeatother

\begin{document}
\title{Moments of Moments and Branching Random Walks}
\abstract
We calculate, for a branching random walk $X_n(l)$ to a leaf $l$ at depth $n$ on a binary tree, the positive integer moments of the random variable $\frac{1}{2^{n}}\sum_{l=1}^{2^n}e^{2\beta X_n(l)}$, for $\beta\in\mathbb{R}$. We obtain explicit formulae for the first few moments for finite $n$. In the limit $n\to\infty$, our expression coincides with recent conjectures and results concerning the moments of moments of characteristic polynomials of random unitary matrices, supporting the idea that these two problems, which both fall into the class of logarithmically correlated Gaussian random fields, are related to each other.

\endabstract

\author{E. C. Bailey}
\email{e.c.bailey@bristol.ac.uk}
\address{School of Mathematics, University of Bristol, Bristol, BS8 1UG, United Kingdom}

\author{J. P. Keating}
\email{keating@maths.ox.ac.uk}
\address{Mathematical Institute, University of Oxford, Oxford, OX2 6GG, United Kingdom}

\maketitle
\section{Introduction}\label{sec:intro}

\subsection{Moments of moments: characteristic polynomials of random matrices}\label{sec:intro1}

In recent years there has been significant progress towards understanding the value distribution of the maximum of the logarithm of the characteristic polynomial of a random unitary matrix and of related log-correlated processes~\cite{arg16, har13a, argbelhar17, argbelbou17, paqzei17, chhmadnaj18, naj18, abbrs19, har19b, argbourad20, argouirad19, web15, niksakweb20, fyohiakea12, fyokea14, fyognukea18, rem20, fyobou08}.  Let
\begin{equation}
  P_N(A,\theta)\coloneqq \det(I-Ae^{-i\theta})
\end{equation}
denote the characteristic polynomial of $A\in U(N)$. Additionally, denote by
\begin{equation}
  P_{\max}(A)\coloneqq \max_{\theta\in[0,2\pi)}\log|P_N(A,\theta)|
\end{equation}
the maximum value of $P_N(A,\theta)$ around the unit circle.  It was conjectured in~\cite{fyohiakea12, fyokea14} that
\begin{equation}\label{eq:max_charpol}
  P_{\max}(A)=\log N-\frac{3}{4}\log\log N+m_N(A)
\end{equation}
where the law of the fluctuating term $m_N(A)$ was postulated to be the same as that of the sum of two independent Gumbel random variables in the limit $N\to\infty$. The leading order of \eqref{eq:max_charpol} was verified by Arguin et al.~\cite{argbelbou17}, and Paquette and Zeitouni~\cite{paqzei17} determined \eqref{eq:max_charpol} to subleading order. At the time of writing, the strongest result in the literature is due to Chhaibi et al.~\cite{chhmadnaj18}, who proved tightness\footnote{In fact, their analysis is more general in that they prove results for the C$\beta$E ensembles, as well as considering the maximum of the imaginary part of $P_N(A,\theta)$.}  of the family of random variables\footnote{Where $N\in\mathbb{N}$.}
\begin{equation}
  \{P_{\max}(N)-\log N+\frac{3}{4}\log\log N\}.
\end{equation}

The maximum conjecture~\eqref{eq:max_charpol} was motivated by a heuristic analysis in~\cite{fyokea14} of the random variable
\begin{equation}\label{eq:circle_av}
  Z_N(A,\theta)\coloneqq \frac{1}{2\pi}\int_0^{2\pi} |P_N(A,\theta)|^{2\beta}d\theta,
\end{equation}
the $2\beta$th moment of the absolute value of the characteristic polynomial with respect to the uniform measure on the unit circle.  In particular, determining the moments of $Z_N(A,\theta)$ with respect to Haar measure on the unitary group is central to the analysis, and such an average is referred to as representing the {\it moments of moments} of $P_N(A,\theta)$.  Specifically, the moments of moments are defined by
\begin{equation}\label{eq:momdef}
  \mom_{U(N)}(k,\beta)\coloneqq \mathbb{E}_{A\in U(N)}\left[\left(\frac{1}{2\pi}\int_0^{2\pi}|P_{N}(A,\theta)|^{2\beta}d\theta\right)^k\right],
\end{equation}
where the external average $\mathbb{E}[\cdot]$ is with respect to the Haar measure on $U(N)$\footnote{One can more generally consider moments of moments of other compact random matrix groups, see for example~\cite{assbaikea20}.}. In~\cite{fyokea14} it was conjectured that, as $N\rightarrow\infty$, $\mom_{U(N)}(k,\beta)$ is given asymptotically by
\begin{equation}\label{eq:fk_conj_mom}
  \mom_{U(N)}(k,\beta)\sim
  \begin{cases}
    \left(\frac{G^2(1+\beta)}{G(1+2\beta)\Gamma(1-\beta^2)}\right)^k\Gamma(1-k\beta^2)N^{k\beta^2},&\text{if }k<1/\beta^2,\\
    c(k,\beta)N^{k^2\beta^2-k+1},&\text{if }k>1/\beta^2,
  \end{cases}
\end{equation}
where $G(s)$ is the Barnes $G$-function, and $c(k,\beta)$ is some (unspecified)  function of the moment parameters $k, \beta$. 

For integer $k, \beta$, it was proved in \cite{baikea19} that $\mom_{U(N)}(k,\beta)$ is a polynomial in the matrix size, $N$, of degree $k^2\beta^2-k+1$, in line with \eqref{eq:fk_conj_mom}.

Using a Riemann-Hilbert analysis, Claeys and Krasovsky~\cite{clakra15} computed $\mom_{U(N)}(2,\beta)$ for $\RE(\beta)>-1/4$, and connected $c(2,\beta)$ to a solution of a Painlev\'e equation. By so doing, they verified \eqref{eq:fk_conj_mom} for $k=2$ and all $\RE(\beta)>-1/4$. Fahs~\cite{fah19} subsequently extended this approach\footnote{A more precise formulation for the leading order coefficient $c(k,\beta)$ in the case $k\geq 3$ is is required in order to draw stronger conclusions regarding $P_{\max}(A)$.} to general $k\in\mathbb{N}$, although he did not determine $c(k,\beta)$ for $k>2$. Additionally, Claeys and Krasovsky, and Fahs, also determined that the behaviour at the critical point $k\beta^2=1$ (still for $k\in\mathbb{N}$) is of the form
\begin{equation}\label{eq:critical_point}
  \mom_{U(N)}(k,\tfrac{1}{\sqrt{k}})\sim\alpha(k,\beta)N\log N,
\end{equation}
for some positive coefficient $\alpha(k,\beta)$ as $N\rightarrow\infty$ (see \cite{clakra15, fah19} for further details). 

One of the key ideas that underpins much of the progress outlined above is that the Fourier series representing $\log P_N(A,\theta)$ exhibits a hierarchical structure typical of problems associated with logarithmically correlated Gaussian fields.  This structure is exemplified by the {\em branching random walk}.  Understanding this connection is currently a focus of research in the area.  Our aim here is to examine it in the context of the moments of moments by calculating the quantity in the theory of the  branching random walk that is analogous to \eqref{eq:momdef}.  Specifically, we will show that the analogue of the moments of moments for the  branching random walk is asymptotically described by a formula that is the direct analogue of \eqref{eq:fk_conj_mom}. Additionally, the fact that $\log P_N(A,\theta)$ has a central limit theorem~\cite{keasna00a} for large $N$ is important to our analysis.

We also remark in passing that the characteristic polynomials of random unitary matrices play an important role in modelling the value distribution of the Riemann zeta-function on its critical line \cite{keasna00a}. There are analogues of the conjectures \eqref{eq:max_charpol} and \eqref{eq:fk_conj_mom} for the zeta function \cite{fyohiakea12, fyokea14}.  In the latter case, the integer moments of moments can be calculated using the shifted moment conjecture of \cite{cfkrs03, cfkrs05}; see \cite{baikea20}.   There has again been a good deal of progress in proving the conjecture corresponding to \eqref{eq:max_charpol} using the analogue for the zeta function of the hierarchical structure exemplified by the branching random walk~\cite{arg16, har13a, argbelhar17, argbelbou17, argouirad19, naj18, abbrs19, har19b, argbourad20}, and so we see our results for the branching random walk as being of interest in the number theoretical context as well.

\subsection{Moments of moments: the branching random walk}\label{branching_unitary}

\begin{figure}[htb]
  \centering
  \begin{tikzpicture}[level distance=1.5cm,
      level 1/.style={sibling distance=4cm},
      level 2/.style={sibling distance=2cm},
      level 3/.style={sibling distance=1cm},
      level 4/.style={sibling distance=.5cm},
      every node/.style = {shape=circle, inner sep=1.5pt,
        draw, align=center,
        fill=black}]
    \node[draw=none,fill=none] at (1.4,-.3) {$Y_1(l)$};
    \node[draw=none,fill=none] at (3.4,-2.2) {$Y_2(l)$};
    \node[draw=none,fill=none] at (4.2,-3.8) {$Y_3(l)$};
    \node[draw=none,fill=none] at (4.5,-5.2) {$Y_4(l)$};
    \node[draw=none,fill=none] at (3.8,-6.4) {$l$};
    \node {}
    child {node {}
      child {node {}
        child {node {}
          child {node {}}
          child {node {}}
        }
        child {node {}
          child {node {}}
          child {node {}}
        }
      }
      child {node {}
        child {node {}
          child {node {}}
          child {node {}}
        }
        child {node {}
          child {node {}}
          child {node {}}
        }
      }
    }
    child {node {}
      child {node {}
        child {node {}
          child {node {}}
          child {node {}}
        }
        child {node {}
          child {node {}}
          child {node {}}
        }
      }
      child {node {}
        child {node {}
          child {node {}}
          child {node {}}
        }
        child {node {}
          child {node {}}
          child {node[fill=white] {}}
        }
      }
    };
  \end{tikzpicture}
  \caption[Example of a random walk on a binary tree.]{An example of a random walk $X_4(l)=Y_1(l)+\cdots+Y_4(l)$ on a binary tree of depth $n=4$, from root to leaf $l$.  The weightings $Y_j(l)$ are independent, centred Gaussian random variables with variance $\frac{1}{2}\log 2$.}\label{fig:binary_tree}
\end{figure}
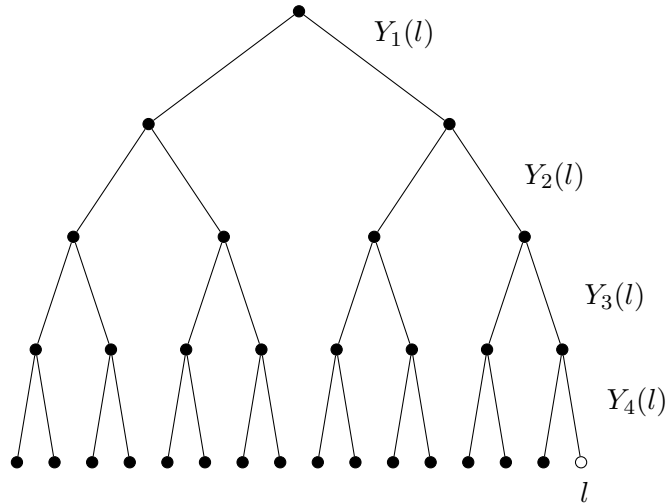

Take a binary tree of depth $n$, and a choice of leaf $l$. Load to each branch in the tree an independent centred Gaussian random variable with variance $\frac{1}{2}\log2$.  We write for the branching random walk from root to $l$
\begin{equation}\label{eq:brw_setup}
  X_n(l)\coloneqq\sum_{m=1}^nY_m(l),
\end{equation}
where $Y_m(l)\sim \cN(0,\frac{1}{2}\log2)$ are the branch weightings, see figure~\ref{fig:binary_tree}.  Note that
\begin{equation}
  X_n(l)\sim \cN\left(0,\tfrac{n}{2}\log 2\right)
\end{equation}
and that the distribution of $X_n(l)$ does not depend on the choice of leaf $l$ (nor does the distribution of $Y_m(l)$ depend on the level $m$ nor the leaf $l$), however including both labels will become useful later. Similarly, it will be important to record the points at which concurrent paths through the tree diverge.

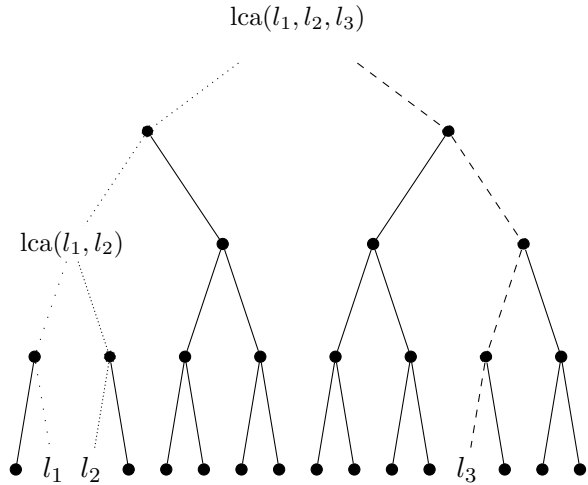
\begin{figure}[htb]
  \centering
  \begin{tikzpicture}[level distance=1.5cm,
      level 1/.style={sibling distance=4cm},
      level 2/.style={sibling distance=2cm},
      level 3/.style={sibling distance=1cm},
      level 4/.style={sibling distance=.5cm},
      every node/.style = {shape=circle, inner sep=1.5pt,
        draw, align=center,
        fill=black}]
    \node[fill=none, draw=none] {\small{$\lca(l_1,l_2,l_3)$}}
    child {[dotted]node {}
      child {[dotted] node[fill=white, draw=none,shape=rectangle] {\small{$\lca(l_1,l_2)$}}
        child {[loosely dotted]node[shape=circle, inner sep=1.5pt, draw, align=center, solid] {}
          child {[solid]node {}}
          child {[loosely dotted]node[fill=white, draw=none] {$l_1$}}
        }
        child {[densely dotted]node {}
          child {[densely dotted]node[fill=white, draw=none] {$l_2$}}
          child {[solid]node {}}
        }
      }
      child {[solid]node {}
        child {node {}
          child {node {}}
          child {node {}}
        }
        child {node {}
          child {node {}}
          child {node {}}
        }
      }
    }
    child {[dashed]node {}
      child {[solid]node {}
        child {node {}
          child {node {}}
          child {node {}}
        }
        child {node {}
          child {node {}}
          child {node {}}
        }
      }
      child {[dashed]node {}
        child {node {}
          child {node[draw=none, fill=white] {$l_3$}}
          child {[solid]node {}}
        }
        child {[solid]node {}
          child {node {}}
          child {node {}}
        }
      }
    };
  \end{tikzpicture}
  \caption[An example of last common ancestors of leaves of a binary tree.]{A binary tree of depth $4$ with three leaves $l_1,l_2,l_3$ highlighted.  The last common ancestor of $l_1,l_2$ is $\lca(l_1,l_2)$.  The last common ancestor of all three (and also $\lca(l_2,l_3)$ and $\lca(l_1,l_3)$) is the root node. The paths are differentiated by dashed and dotted lines.}\label{fig:lca}
\end{figure}

\begin{definition}
  Take two leaves $l_1, l_2$ of a binary tree of depth $n$.  The \emph{last common ancestor} of $l_1, l_2$, denoted by $\lca(l_1,l_2)$ is the furthest node from the root that has both $l_1$ and $l_2$ as descendants.  The last common ancestor of $k$ leaves is the furthest node from the root with all $k$ leaves as descendants. Figure~\ref{fig:lca} shows an example involving three leaves on a tree of depth $n=4$.
  
It will be important for our purposes to keep track of the level of the last common ancestor.  Hence, we also define the \emph{last common level} $\lcl(l_1,\dots,l_k)$ to be the level of $\lca(l_1,\dots,l_k)$. For example, in figure~\ref{fig:lca}, $\lcl(l_1,l_2,l_3)=0$ and $\lcl(l_1,l_2)=2$.  
\end{definition}

As a process, $\{X_n(l), l\in\{1,\dots 2^n\}\}$ is log-correlated (see for example~\cite{arg16}).  It is natural, therefore, to investigate the associated partition function\footnote{In \eqref{eq:final_branching_model}, the `temperature' parameter is $2\beta$ rather than $-\beta$, so as to be in keeping with the random matrix literature.} or moment generating function
\begin{align}
  \frac{1}{2^{n}}\sum_{l=1}^{2^n}e^{2\beta X_n(l)}&=\frac{1}{2^{n}}\sum_{l=1}^{2^n}e^{2\beta \sum_{m=1}^nY_m(l)}\label{eq:final_branching_model}
\end{align}
where, as in \eqref{eq:brw_setup}, $Y_m(l)\sim\cN(0,\frac{1}{2}\log 2)$ and are independent. 

In particular, we are interested in the moments of the partition function \eqref{eq:final_branching_model}, 
\begin{align}
  \mathbb{E}\left[\left(\frac{1}{2^n}\sum_{l=1}^{2^n}e^{2\beta X_n(l)}\right)^k\right]&=\frac{1}{2^{kn}}\sum_{l_1=1}^{2^n}\cdots\sum_{l_k=1}^{2^n}\mathbb{E}\left[e^{2\beta(X_n(l_1)+\cdots+X_n(l_k))}\right],\label{branching_model_mom}
\end{align}
where the expectation in \eqref{branching_model_mom} is with respect to the Gaussian random variables.  These are the \emph{moments of moments} for the branching random walk.  They are the analogues of \eqref{eq:momdef}.

\section{Results and Proof Outline}\label{sec:results_and_outline}

As reviewed in section~\ref{sec:intro}, it is now known that (see~\cite{baikea19, asskea20, clakra15, fah19}) for $\beta\geq 0$, and $k\in\mathbb{N}$ 
\begin{equation}\label{branching_restated_mom}
  \mom_{U(N)}(k,\beta)\sim 
  \begin{cases}
    \left(\frac{G^2(1+\beta)}{G(1+2\beta)\Gamma(1-\beta^2)}\right)^k\Gamma(1-k\beta^2)N^{k\beta^2},&\text{if }k<1/\beta^2,\\
    \alpha(k,\beta)N\log N,&\text{if }k=1/\beta^2,\\
    c(k,\beta)N^{k^2\beta^2-k+1},&\text{if }k>1/\beta^2,
  \end{cases}
\end{equation}
as $N\rightarrow\infty$ for some positive constants\footnote{For further details on the form of $\alpha(k,\beta)$ and  $c(k,\beta)$ see~\cite{clakra15} for the case $k=2$, and~\cite{asskea20, baikea19} for expressions for $c(k,\beta)$ for $k\geq 3$.} $\alpha(k,\beta)$ and $c(k,\beta)$ depending only on $k,\beta$.  Furthermore, for $k,\beta\in\mathbb{N}$, $\mom_{U(N)}(k,\beta)$ is a polynomial in $N$, see~\cite{baikea19}.

\subsection{Results}\label{sec:results}
By calculating the moments of moments \eqref{branching_model_mom}, we are able to recover an asymptotic result of the form~\eqref{branching_restated_mom}, albeit with different leading order coefficients. Explicitly, we prove the following.
\begin{theorem}\label{thm:branching_mom}
  Take $n, k\in\mathbb{N}$ and $\beta\in\mathbb{R}$. If $\beta\neq0$ then
  \begin{equation}\label{eq:branching_asympt}
    \mathbb{E}\left[\left(\frac{1}{2^n}\sum_{l=1}^{2^n}e^{2\beta X_n(l)}\right)^k\right]\sim
    \begin{cases}
      \rho(k,\beta)2^{k\beta^2n},&\text{if }k<1/\beta^2,\\
      \sigma(k,\beta)n2^n,&\text{if }k=1/\beta^2,\\
      \tau(k,\beta)2^{(k^2\beta^2-k+1)n},&\text{if }k>1/\beta^2,
    \end{cases}
  \end{equation}
  as $n\rightarrow\infty$, for some positive constants $\rho(k,\beta), \sigma(k,\beta),$ and $\tau(k,\beta)$ depending only on $k,\beta$.  Clearly, if $\beta=0$ then the expectation evaluates to $1$. 
\end{theorem}

For small values of $k$, one can calculate exact and explicit formulae for the moments of moments; we provide such examples for $k=1,\dots,5$ in appendix~\ref{sec:appendix}. Such moments were also considered by Derrida and Spohn~\cite{derspo88} for general branching weightings and in continuous settings.  In the discrete setting, they compute the first few low moments (which agree with the first few explicit examples computed in appendix~\ref{sec:appendix}) and establish a connection to the KPP equation in the continuous setting.  Additionally, it is natural to ask if the leading order coefficient $\rho(k,\beta)$ for $k\beta^2<1$ in \eqref{eq:branching_asympt} could take the form $f(\beta)^k\Gamma(1-k\beta^2)$ for some function $f$, in line with Fyodorov-Bouchaud~\cite{fyobou08} or Remy-Zhu~\cite{remzhu20} formulae for related problems in the same regime (see also~\eqref{branching_restated_mom}).  Such a statement does not appear to hold here. 

Furthermore, we are able to establish that for integer values of the moment parameters the branching moments of moments are polynomials. 

\begin{corollary}\label{cor:branching_mom}
When $k,\beta\in\mathbb{N}$, \eqref{branching_model_mom} is a polynomial in $2^n$ of degree $k^2\beta^2-k+1$. 
\end{corollary}
Thus, the branching moments of moments exhibit asymptotic behaviour identical to that of the random matrix moments of moments, once the identification $N=2^n$ is made. 

The remainder of this section details the key ideas necessary for the proof of theorem~\ref{thm:branching_mom} and corollary~\ref{cor:branching_mom}.  Small cases of the moments of moments are explicitly calculated.

\subsection{Structure of proof}\label{sec:structure}

Establishing the statement of theorem~\ref{thm:branching_mom} in the simplest instance, $k=1$, follows from a moment generating function calculation. Recall that in the random matrix case, $\mom_{U(N)}(1,\beta)$ has an exact (finite $N$) expression:
\begin{equation}\label{unitary_mom_k1}
  \mom_{U(N)}(1,\beta)=\prod_{j=1}^N\frac{\Gamma(j+2\beta)\Gamma(j)}{\Gamma^2(j+\beta)}
\end{equation}
for $\RE(\beta)>-\frac{1}{2}$, see~\cite{keasna00a}. For $\beta\in\mathbb{N}$, the right hand side of \eqref{unitary_mom_k1} simplifies to
\begin{equation}\label{eq:unitary_mom_int}
  \prod_{0\leq i, j\leq \beta-1}\left(\frac{N}{i+j+1}+1\right).
\end{equation}
As $N\rightarrow\infty$,
\begin{equation}\label{eq:unitary_mom_k1}
  \mom_{U(N)}(1,\beta)\sim c(1,\beta)N^{\beta^2}
\end{equation}
where $c(1,\beta)$ is the ratio of Barnes $G$-functions appearing in the first regime in \eqref{branching_restated_mom}.  The asymptotic behaviour~\eqref{eq:unitary_mom_k1} for integer $\beta$ follows from~\eqref{eq:unitary_mom_int}; for general $\beta$ it was determined by Keating and Snaith~\cite{keasna00a}. As is consistent with \eqref{branching_restated_mom}, for $k=1$ there is no phase transition as $\beta$ varies.   

The equivalent case of $k=1$ for the branching moments of moments (see~\eqref{branching_model_mom}) requires calculating the following moment
\begin{equation}\label{eq:k=1a}
  \frac{1}{2^n}\mathbb{E}\left[\sum_{l=1}^{2^n}e^{2\beta X_n(l)}\right]=\frac{1}{2^n}\sum_{l=1}^{2^n}\mathbb{E}\left[\prod_{j=1}^ne^{2\beta Y_j(l)}\right].
\end{equation}
In terms of the binary tree, this can be interpreted as `loading' the root with one particle.  Consequently, each summand is the contribution from that particle passing through the tree and ending at leaf $l$. Since the $Y_j(l)$ are independent between each level of the binary tree, we have
\begin{equation}\label{eq:k=1b}
  \frac{1}{2^n}\sum_{l=1}^{2^n}\mathbb{E}\left[\prod_{j=1}^ne^{2\beta Y_j(l)}\right]=\frac{1}{2^n}\sum_{l=1}^{2^n}\prod_{j=1}^n\mathbb{E}\left[e^{2\beta Y_j(l)}\right]=2^{\beta^2n}
\end{equation}
since $Y_j(l)\sim\cN(0,\frac{1}{2}\log 2)$. By making the identification $N=2^n$, the branching moments of moments exhibit the same asymptotic growth (although with a different leading order coefficient, and no lower order terms) as \eqref{eq:unitary_mom_k1}.

When $k\geq 2$, one has the additional difficulty of the paths $X_n(l_j)$ no longer being independent typically.  In order to introduce the key ideas of the proof for general $k$, it is instructive also to calculate explicitly the case for $k=2$. This case is the first where a phase change can be seen as $\beta$ varies, and the calculation demonstrates how to handle the dependence between paths. For ease of notation, henceforth we write for the branching moments of moments in \eqref{branching_model_mom}
\begin{equation}
  \mom_n(k,\beta)\coloneqq \frac{1}{2^{kn}}\sum_{l_1=1}^{2^n}\cdots \sum_{l_k=1}^{2^n}\mathbb{E}\left[e^{2\beta(X_n(l_1)+\cdots+X_n(l_k))}\right].\label{eq:mom}
\end{equation}

Additionally, since the case for $\beta=0$ is trivial, henceforth we assume $\beta\neq 0$. Thus, take $\beta\neq0$ and consider \eqref{eq:mom} for $k=2$, 
\begin{align}
  \mom_n(2,\beta)&=\frac{1}{2^{2n}}\sum_{l_1=1}^{2^n}\sum_{l_2=1}^{2^n}\mathbb{E}\left[e^{2\beta(X_n(l_1)+X_n(l_2))}\right]\\
  &=\frac{1}{2^{2n}}\sum_{l_1=1}^{2^n}\sum_{l_2=1}^{2^n}\mathbb{E}\left[\prod_{j=1}^\lambda e^{2\beta(Y_j(l_1)+Y_j(l_2))}\right]\mathbb{E}\left[\prod_{j=\lambda+1}^n e^{2\beta(Y_j(l_1)+Y_j(l_2))}\right]\label{eq:k2_inductive0}
\end{align}
where $\lambda\coloneqq\lcl(l_1,l_2)$. As up to level $\lambda$ the paths are identical, and thereafter independent, we may rewrite \eqref{eq:k2_inductive0} as  
\begin{equation}
  \frac{1}{2^{2n}}\left(\sum_{\lambda=0}^{n-1}2^{\lambda}2^{4\beta^2\lambda}\mathbb{E}\left[\prod_{j=\lambda+1}^n e^{2\beta Y_{j}}\right]^2+2^{(4\beta^2+1)n}\right).\label{eq:k2_inductive1}
\end{equation}
This follows because $2^{4\beta\lambda}$ is the contribution from the joined paths, and $2^\lambda$ is the number of choices of $\lca(l_1,l_2)$ given $\lcl(l_1,l_2)=\lambda$. 

At this point observe that the expectation on the right hand side of \eqref{eq:k2_inductive1} is the same as calculated for the first moment of moments, except on a tree of depth $n-\lambda-1$ (and with an additional step prior to the new root node). Hence we proceed inductively,
\begin{align}
  \mom_n(2,\beta)&=\frac{1}{2^{2n}}\Bigg(\sum_{\lambda=0}^{n-1}2^{\lambda}2^{4\beta^2\lambda}\binom{2}{1}\left(2^{\beta^2}2^{n-\lambda-1}\mom_{n-\lambda-1}(1,\beta)\right)^2+2^{(4\beta^2+1)n}\Bigg)\nonumber\\
  &=2^{2\beta^2-1}\sum_{\lambda=0}^{n-1}2^{(4\beta^2-1)\lambda}2^{2\beta^2(n-\lambda-1)}+2^{(4\beta^2-1)n}\label{eq:k=2_interim}\\
  &=2^{2\beta^2n-1}\frac{2^{(2\beta^2-1)n}-1}{2^{2\beta^2-1}-1}+2^{(4\beta^2-1)n}.\label{eq:k2_sum}
\end{align}
Thus, the general method for proving theorem~\ref{thm:branching_mom}, and hence corollary~\ref{cor:branching_mom}, will follow via strong induction. In order to demonstrate the three different asymptotic regimes, we examine \eqref{eq:k2_sum} for different values of $\beta$.  

If $2\beta^2>1$ then
\begin{align}
  \mom_n(2,\beta)&\sim \left(1+\frac{1}{2(2^{2\beta^2-1}-1)}\right)2^{(4\beta^2-1)n}\\
  &=\frac{2^{2\beta^2}-1}{2(2^{2\beta^2-1}-1)}2^{(4\beta^2-1)n},\label{eq:mom2_1}
\end{align}
as $n\rightarrow\infty$. 

Instead, if $2\beta^2<1$, then as $n\rightarrow\infty$
\begin{equation}\label{eq:mom2_2}
  \mom_n(2,\beta)\sim \frac{1}{2(1-2^{2\beta^2-1})}2^{2\beta^2n}.
\end{equation}

Finally, if $2\beta^2=1$, then using \eqref{eq:k=2_interim} we have
\begin{align}
  \mom_n\left(2,\tfrac{1}{\sqrt{2}}\right)&=\lim_{2\beta^2\rightarrow 1}\left(2^{2\beta^2n}\sum_{\lambda=0}^{n-1}(2^{(2\beta^2-1)\lambda}-2^{(2\beta^2-1)\lambda-1})+2^{(4\beta^2-1)n}\right)\\
  &=\frac{n+2}{2}2^n.\label{eq:mom2_3}
\end{align}
Hence, as $n\rightarrow\infty$, at $2\beta^2=1$,
\begin{equation}
  \mom_n\left(2,\tfrac{1}{\sqrt{2}}\right)\sim \frac{n}{2}2^n.
\end{equation}

In the next section, we prove theorem~\ref{thm:branching_mom} and corollary~\ref{cor:branching_mom} using the techniques presented in this section.  In particular, we make liberal use of the iterative properties of the binary tree underpinning \eqref{eq:mom}.

\section{Proof details}

We proceed by strong induction. Recall that we write for $\beta\in\mathbb{R}$ and $k\in\mathbb{N}$
\begin{equation}
  \mom_n(k,\beta)=\frac{1}{2^{kn}}\sum_{l_1=1}^{2^n}\cdots\sum_{l_k=1}^{2^n}\mathbb{E}\left[e^{2\beta(X_n(l_1)+\cdots+X_n(l_k))}\right].
\end{equation}
In section~\ref{sec:structure} we established the base cases of $\mom_n(1,\beta), \mom_n(2,\beta)$. 
As the case $\beta=0$ is trivial, here and henceforth $\beta\neq 0$. Now assume for all  $j< k$, and $k\geq 2$, that

\begin{align}\label{eq:induction_hyp}
  \mom_n(j,\beta)&\sim
  \begin{cases}
    \rho(j,\beta)2^{j\beta^2n},&\text{if }j\beta^2<1,\\
    \sigma(j,\beta)n2^n,&\text{if }j\beta^2=1,\\
    \tau(j,\beta)2^{(j^2\beta^2-j+1)n},&\text{if }j\beta^2>1,
  \end{cases}
\end{align}
where $\rho(j,\beta),\sigma(j,\beta),\tau(j,\beta)$ are the positive leading order coefficients (depending on the moment parameters $j$ and $\beta$) of $\mom_n(j,\beta)$ in each of the three regimes\footnote{For example, $\rho(2,\beta)$, $\sigma(2,\beta)$, and $\tau(2,\beta)$ are given respectively by \eqref{eq:mom2_1}, \eqref{eq:mom2_2}, and \eqref{eq:mom2_3}.  Although the $k=1$ case exhibits no phase transition, we will write for ease of notation $\rho(1,\beta)\equiv\sigma(1,\beta)\equiv\tau(1,\beta)=1$.}.

Throughout we write $\Sigma^\prime$ for a sum without the diagonal term.  We now consider the $k$th case,
\begin{align}
  \mom_n(k,\beta)
  &=\frac{1}{2^{kn}}\sum_{l_1=1}^{2^n}\cdots\sum_{l_k=1}^{2^n}\mathbb{E}\left[e^{2\beta(X_n(l_1)+\cdots+X_n(l_k))}\right]\\
  &=\frac{1}{2^{kn}}\left[\sideset{}{^\prime}\sum_{l_1,\dots,l_k=1}^{2^n}\mathbb{E}\left[e^{2\beta(X_n(l_1)+\cdots+X_n(l_k))}\right]+2^{(k^2\beta^2+1)n}\right]\\
  &=\frac{1}{2^{kn}}\sideset{}{^\prime}\sum_{l_1,\dots,l_k=1}^{2^n}\mathbb{E}\left[\prod_{j=1}^k\prod_{i=1}^{\lambda} e^{2\beta Y_i(l_j)}\right]\mathbb{E}\left[\prod_{j=1}^k\prod_{i=\lambda+1}^{n}e^{2\beta Y_i(l_j)}\right]+2^{(k^2\beta^2-k+1)n}\\
  &=\frac{1}{2^{kn}}\sideset{}{^\prime}\sum_{l_1,\dots,l_k=1}^{2^n}2^{ k^2\beta^2\lambda}\mathbb{E}\left[\prod_{j=1}^ke^{2\beta(Y_{\lambda+1}(l_j)+\cdots+Y_n(l_j))}\right]+2^{(k^2\beta^2-k+1)n},
\end{align}
where in the last two lines $\lambda\coloneqq\lcl(l_1,\dots,l_k)$. At the initial separation on level $\lambda$, $j$ particles will split in one direction, and $k-j$ in the other for $j\in\{1,\dots,k-1\}$.  Thereafter, one is essentially the dealing with two subtrees of depth $n-\lambda-1$, with $j$ particles on one and $k-j$ on the other.  Note also that there are $2^\lambda$ choices for $\lca(l_1,\dots,l_k)$ given $\lcl(l_1,\dots,l_k)=\lambda$, and that since only off-diagonal terms appear in the sum, $\lambda\in\{0,\dots,n-1\}$.  Let $Y\sim\cN(0,\frac{1}{2}\log 2)$, then
\begin{align}
  \mom_n(k,\beta)&=\frac{1}{2^{kn}}\sum_{\lambda=0}^{n-1}2^{(k^2\beta^2+1)\lambda}\sum_{j=1}^{k-1}\binom{k}{j}\mathbb{E}\left[e^{2\beta j Y}\right]\mathbb{E}\left[e^{2\beta(k-j) Y}\right]\nonumber\\
  &\qquad\quad\times\Big(2^{j(n-\lambda-1)}\mom_{n-\lambda-1}(j,\beta)\Big)\Big(2^{(k-j)(n-\lambda-1)}\mom_{n-\lambda-1}(k-j,\beta)\Big)\nonumber\\
  &\quad+2^{(k^2\beta^2-k+1)n}\\
  &=\frac{1}{2^{kn}}\sum_{\lambda=0}^{n-1}2^{(k^2\beta^2+1)\lambda}\sum_{j=1}^{k-1}\binom{k}{j}2^{\beta^2j^2}2^{\beta^2(k-j)^2}2^{k(n-\lambda-1)}\nonumber\\
    &\qquad\quad\times \mom_{n-\lambda-1}(j,\beta)\mom_{n-\lambda-1}(k-j,\beta)+2^{(k^2\beta^2-k+1)n}\\
  &=2^{k^2\beta^2-k}\sum_{\lambda=0}^{n-1}2^{(k^2\beta^2-k+1)\lambda}\sum_{j=1}^{k-1}\binom{k}{j}2^{2j\beta^2(j-k)}\nonumber\\
    &\qquad\quad\times \mom_{n-\lambda-1}(j,\beta)\mom_{n-\lambda-1}(k-j,\beta)+2^{(k^2\beta^2-k+1)n}.\label{eq:inductive_decomp1}
\end{align}

To complete the proof of theorem~\ref{thm:branching_mom}, we determine the asymptotic behaviour of \eqref{eq:inductive_decomp1} by separately considering the ranges $k\beta^2<1$, $k\beta^2=1$, and $k\beta^2>1$.  It transpires that we will need to further partition the case $k\beta^2>1$, for more details see section~\ref{sec:third_regime}. 

\subsection{Range: $0<|\beta|<\frac{1}{\sqrt{k}}$}\label{sec:first_regime}
In this range $k\beta^2<1$ so we expect $\mom_n(k,\beta)$ to grow as $2^{k\beta^2n}$.  Further, since $k\beta^2<1$, we also have $m\beta^2<1$ for $m=1,\dots,k-1$. From \eqref{eq:inductive_decomp1}, we have that
\begin{align}
  \mom_n(k,\beta)&=2^{k^2\beta^2-k}\sum_{\lambda=0}^{n-1}2^{(k^2\beta^2-k+1)\lambda}\sum_{j=1}^{k-1}\binom{k}{j}2^{2j\beta^2(j-k)}\nonumber\\
  &\qquad\quad\times \mom_{n-\lambda-1}(j,\beta)\mom_{n-\lambda-1}(k-j,\beta)+2^{(k^2\beta^2-k+1)n}\\
    &\sim 2^{k^2\beta^2-k}\sum_{\lambda=0}^{n-1}2^{(k^2\beta^2-k+1)\lambda}\sum_{j=1}^{k-1}\binom{k}{j}2^{2j\beta^2(j-k)}\nonumber\\
    &\qquad\quad\times \rho(j,\beta)\rho(k-j,\beta)2^{j\beta^2(n-\lambda-1)}2^{(k-j)\beta^2(n-\lambda-1)}+2^{(k^2\beta^2-k+1)n}\\
  &=2^{k^2\beta^2-k}2^{k\beta^2(n-1)}\sum_{j=1}^{k-1}\binom{k}{j}2^{2j\beta^2(j-k)}\rho(j,\beta)\rho(k-j,\beta)\sum_{\lambda=0}^{n-1}2^{(k^2\beta^2-k+1-k\beta^2)\lambda}\nonumber\\
  &\qquad +2^{(k^2\beta^2-k+1)n}\\
  &=2^{k\beta^2(n-1)}\frac{2^{(k^2\beta^2-k+1-k\beta^2)n}-1}{2^{k^2\beta^2-k+1-k\beta^2}-1}\left[2^{k^2\beta^2-k}\sum_{j=1}^{k-1}\binom{k}{j}2^{2j\beta^2(j-k)}\rho(j,\beta)\rho(k-j,\beta)\right]\nonumber\\
  &\qquad+2^{(k^2\beta^2-k+1)n}.
\end{align}
Define
\begin{equation}
  \pi(k,\beta)\coloneqq2^{k^2\beta^2-k}\sum_{j=1}^{k-1}\binom{k}{j}2^{2j\beta^2(j-k)}\rho(j,\beta)\rho(k-j,\beta).
\end{equation}
Hence, for $0<\abs{\beta}<\frac{1}{\sqrt{k}}$,
\begin{align}
  \mom_n(k,\beta)&\sim \pi(k,\beta)2^{k\beta^2(n-1)}\frac{2^{(k^2\beta^2-k+1-k\beta^2)n}-1}{2^{k^2\beta^2-k+1-k\beta^2}-1}+2^{(k^2\beta^2-k+1)n}\\
  &=\pi(k,\beta)\frac{2^{(k\beta^2)n}-2^{(k^2\beta^2-k+1)n}}{2^{k\beta^2}-2^{k^2\beta^2-k+1}}+2^{(k^2\beta^2-k+1)n}.
\end{align}
Observe that since $k\beta^2<1$ and $k\geq 2$,
\begin{align}
  k^2\beta^2+k-1<k\beta^2. 
\end{align}
Hence, for $0<\abs{\beta}<\frac{1}{\sqrt{k}}$, 
\begin{align}
  \mom_n(k,\beta)
  &\sim \rho(k,\beta)2^{k\beta^2n},
\end{align}
as $n\rightarrow\infty$ where $\rho(k,\beta)\coloneqq \pi(k,\beta)(2^{k\beta^2}-2^{k^2\beta^2-k+1})^{-1}$.

\subsection{Range: $|\beta|=\frac{1}{\sqrt{k}}$}\label{sec:second_regime}
For this value of $\beta$, we expect $\mom_n(k,\beta)$ to grow like $n2^n$. Additionally, for $k\beta^2=1$, one has $m\beta^2<1$ for $m=1,\dots,k-1$. From \eqref{eq:inductive_decomp1} we have 
\begin{align}
  \mom_n(k,\beta)&=2^{k^2\beta^2-k}\sum_{\lambda=0}^{n-1}2^{(k^2\beta^2-k+1)\lambda}\sum_{j=1}^{k-1}\binom{k}{j}2^{2j\beta^2(j-k)}\nonumber\\
  &\qquad\quad\times \mom_{n-\lambda-1}(j,\beta)\mom_{n-\lambda-1}(k-j,\beta)+2^{(k^2\beta^2-k+1)n}\\
  &=\sum_{j=1}^{k-1}\binom{k}{j}2^{\frac{2j}{k}(j-k)}\sum_{\lambda=0}^{n-1}2^\lambda\mom_{n-\lambda-1}(j,\beta)\mom_{n-\lambda-1}(k-j,\beta)+2^{n}\\
  &\sim\sum_{j=1}^{k-1}\binom{k}{j}2^{\frac{2j}{k}(j-k)}\rho(j,\beta)\rho(k-j,\beta)\sum_{\lambda=0}^{n-1}2^\lambda2^{(n-\lambda-1)}+2^{n}\\
  &=n2^{n}\sum_{j=1}^{k-1}\binom{k}{j}2^{\frac{2j}{k}(j-k)-1}\rho(j,\beta)\rho(k-j,\beta)+2^{n}.
\end{align}
Hence, as $n\rightarrow\infty$
\begin{equation}
  \mom_n(k,\frac{1}{\sqrt{k}})\sim \sigma(k,\beta)n2^n,
\end{equation}
where 
\begin{equation}
  \sigma(k,\beta)\coloneqq \frac{1}{2}\sum_{j=1}^{k-1}\binom{k}{j}2^{\frac{2j}{k}(j-k)}\rho(j,\beta)\rho(k-j,\beta).
\end{equation}

\subsection{Range: $|\beta|>\frac{1}{\sqrt{k}}$}\label{sec:third_regime}

In this range, $k\beta^2>1$ so we expect $\mom_n(k,\beta)$ to grow like $2^{(k^2\beta^2-k+1)n}$. As previously mentioned, it will be necessary to further partition the span of values.  The three divisions\footnote{Though only when $k\geq 3$ are all three cases required.} are:
\begin{itemize}
\item $\frac{1}{\sqrt{m}}<\abs{\beta}<\frac{1}{\sqrt{m-1}}$, for $m=3,\dots,k$,
\item $\frac{1}{\sqrt{2}}<\abs{\beta}$,
\item $\abs{\beta}=\frac{1}{\sqrt{m}}$, for $m=2,\dots,k-1$.
\end{itemize}

\subsubsection{Range: $\frac{1}{\sqrt{m}}<|\beta|<\frac{1}{\sqrt{m-1}}$}\label{sec:beta_1}
Assume that $\frac{1}{\sqrt{m}}<\abs{\beta}<\frac{1}{\sqrt{m-1}}$ for some $m\in\{3,\dots,k\}$. We first record a useful rewriting of \eqref{eq:inductive_decomp1} due to the symmetric nature of the summands.  If $k$ is odd then
\begin{align}
  \mom_n(k,\beta)&=2^{k^2\beta^2-k+1}\sum_{\lambda=0}^{n-1}2^{(k^2\beta^2-k+1)\lambda}\sum_{j=1}^{\frac{k-1}{2}}\binom{k}{j}2^{2j\beta^2(j-k)}\nonumber\\
    &\qquad\quad\times \mom_{n-\lambda-1}(j,\beta)\mom_{n-\lambda-1}(k-j,\beta)+2^{(k^2\beta^2-k+1)n}.\label{eq:k_odd}
\end{align}

Instead if $k$ is even, then
\begin{align}
  \mom_n(k,\beta)&=2^{k^2\beta^2-k+1}\sum_{\lambda=0}^{n-1}2^{(k^2\beta^2-k+1)\lambda}\sum_{j=1}^{\frac{k-2}{2}}\binom{k}{j}2^{2j\beta^2(j-k)}\nonumber\\
  &\qquad\quad\times \mom_{n-\lambda-1}(j,\beta)\mom_{n-\lambda-1}(k-j,\beta)\nonumber\\
  &\quad +2^{\frac{k^2\beta^2}{2}-k}\binom{k}{\frac{k}{2}}\sum_{\lambda=0}^{n-1}2^{(k^2\beta^2-k+1)\lambda}\left(\mom_{n-\lambda-1}(\tfrac{k}{2},\beta)\right)^2+2^{(k^2\beta^2-k+1)n}.\label{eq:k_even}
\end{align}

In either case, $\mom_n(j,\beta)$ is paired with $\mom_n(k-j,\beta)$. Hence, we first consider the case of $2<m\leq\left\lfloor\frac{k}{2}\right\rfloor$ and split the sums at $m$ in order to apply \eqref{eq:induction_hyp}. Then, \eqref{eq:k_odd} becomes
\begin{align}
  \mom_n(k,\beta)
  &=2^{k^2\beta^2-k+1}\sum_{\lambda=0}^{n-1}2^{(k^2\beta^2-k+1)\lambda}\nonumber\\
  &\qquad\Bigg[\sum_{j=1}^{m-1}\binom{k}{j}2^{2j\beta^2(j-k)}\mom_{n-\lambda-1}(j,\beta)\mom_{n-\lambda-1}(k-j,\beta)\nonumber\\
    &\qquad +\sum_{j=m}^{\frac{k-1}{2}}\binom{k}{j}2^{2j\beta^2(j-k)}\mom_{n-\lambda-1}(j,\beta)\mom_{n-\lambda-1}(k-j,\beta)\Bigg]\nonumber\\
    &\qquad+2^{(k^2\beta^2-k+1)n}.\label{eq:k_odd_first}
\end{align}

Instead if $k$ is even and $2<m\leq\frac{k}{2}$
\begin{align}
  \mom_n(k,\beta)
   &=2^{k^2\beta^2-k+1}\sum_{\lambda=0}^{n-1}2^{(k^2\beta^2-k+1)\lambda}\nonumber\\
  &\qquad\Bigg[\sum_{j=1}^{m-1}\binom{k}{j}2^{2j\beta^2(j-k)}\mom_{n-\lambda-1}(j,\beta)\mom_{n-\lambda-1}(k-j,\beta)\nonumber\\
   &\qquad+\sum_{j=m}^{\frac{k}{2}-1}\binom{k}{j}2^{2j\beta^2(j-k)}\mom_{n-\lambda-1}(j,\beta)\mom_{n-\lambda-1}(k-j,\beta)\Bigg]\nonumber\\
  &\quad +2^{\frac{k^2\beta^2}{2}-k}\binom{k}{\frac{k}{2}}\sum_{\lambda=0}^{n-1}2^{(k^2\beta^2-k+1)\lambda}\left(\mom_{n-\lambda-1}(\tfrac{k}{2},\beta)\right)^2+2^{(k^2\beta^2-k+1)n}.\label{eq:k_even_first}
\end{align}

If $\left\lfloor\frac{k}{2}\right\rfloor<m\leq k$ and $k$ odd,
\begin{align}
  \mom_n(k,\beta)
  &=2^{k^2\beta^2-k+1}\sum_{\lambda=0}^{n-1}2^{(k^2\beta^2-k+1)\lambda}\nonumber\\
  &\qquad\Bigg[\sum_{j=1}^{k-m}\binom{k}{j}2^{2j\beta^2(j-k)}\mom_{n-\lambda-1}(j,\beta)\mom_{n-\lambda-1}(k-j,\beta)\nonumber\\
    &\qquad +\sum_{j=k-m+1}^{\frac{k-1}{2}}\binom{k}{j}2^{2j\beta^2(j-k)}\mom_{n-\lambda-1}(j,\beta)\mom_{n-\lambda-1}(k-j,\beta)\Bigg]\nonumber\\
    &\qquad+2^{(k^2\beta^2-k+1)n}.\label{eq:k_odd_second}
\end{align}

If $\frac{k}{2}<m\leq k$ and $k$ even,
\begin{align}
  \mom_n(k,\beta)
  &=2^{k^2\beta^2-k+1}\sum_{\lambda=0}^{n-1}2^{(k^2\beta^2-k+1)\lambda}\nonumber\\
  &\qquad\Bigg[\sum_{j=1}^{k-m}\binom{k}{j}2^{2j\beta^2(j-k)}\mom_{n-\lambda-1}(j,\beta)\mom_{n-\lambda-1}(k-j,\beta)\nonumber\\
   &\qquad+\sum_{j=k-m+1}^{\frac{k}{2}-1}\binom{k}{j}2^{2j\beta^2(j-k)}\mom_{n-\lambda-1}(j,\beta)\mom_{n-\lambda-1}(k-j,\beta)\Bigg]\nonumber\\
  &\quad +2^{\frac{k^2\beta^2}{2}-k}\binom{k}{\frac{k}{2}}\sum_{\lambda=0}^{n-1}2^{(k^2\beta^2-k+1)\lambda}\left(\mom_{n-\lambda-1}(\tfrac{k}{2},\beta)\right)^2+2^{(k^2\beta^2-k+1)n}.\label{eq:k_even_second}
\end{align}

Now, applying \eqref{eq:induction_hyp} to \eqref{eq:k_odd_first}, for odd $k$ and $2<m\leq\left\lfloor\frac{k}{2}\right\rfloor$
\begin{align}
  \mom_n(k,\beta)
  &\sim 2^{k^2\beta^2-k+1}\sum_{\lambda=0}^{n-1}2^{(k^2\beta^2-k+1)\lambda}\nonumber\\
  &\qquad\Bigg[\sum_{j=1}^{m-1}\binom{k}{j}2^{2j\beta^2(j-k)}\rho(j,\beta)\tau(k-j,\beta)2^{(j\beta^2+(k-j)^2\beta^2-(k-j)+1)(n-\lambda-1)}\nonumber\\
    &\qquad +\sum_{j=m}^{\frac{k-1}{2}}\binom{k}{j}2^{2j\beta^2(j-k)}\tau(j,\beta)\tau(k-j,\beta)2^{(j^2\beta^2-j+1+(k-j)^2\beta^2-(k-j)+1)(n-\lambda-1)}\Bigg]\nonumber\\
  &\qquad+2^{(k^2\beta^2-k+1)n}\\
  &=2^{(k^2\beta^2-k+1)n}\Bigg[\sum_{j=1}^{m-1}\binom{k}{j}\rho(j,\beta)\tau(k-j,\beta)2^{j\beta^2(j-1)-j}\frac{1-2^{j(\beta^2(j+1-2k)+1)n}}{2^{j(\beta^2(2k-j-1)-1)}-1}\nonumber\\
  &\qquad+\sum_{j=m}^{\frac{k-1}{2}}\binom{k}{j}\tau(j,\beta)\tau(k-j,\beta)\frac{1-2^{(2j\beta^2(j-k)+1)n}}{2(2^{2j\beta^2(k-j)-1}-1)}+1\Bigg].
\end{align}
Hence, in order to show that $\mom_n(k,\beta)$ grows like $2^{(k^2\beta^2-k+1)n}$, we need to establish both that $2^{j(\beta^2(j+1-2k)+1)n}$ is subleading, for $j=1,\dots,m-1$, as well as $2^{(2j\beta^2(j-k)+1)n}$ for $j=m,\dots,\frac{k-1}{2}$, provided $\frac{1}{\sqrt{m}}<\abs{\beta}<\frac{1}{\sqrt{m-1}}$ and $2<m\leq\frac{k-1}{2}$. 

In the first case, we have $j\beta^2<1$ for $j=1,\dots,m-1$ as $\frac{1}{\sqrt{m}}<\abs{\beta}<\frac{1}{\sqrt{m-1}}$. Further
\begin{equation}\label{eq:k_odd_first2}
  j\beta^2<1<(2k-1)\beta^2-1
\end{equation}
since $l\beta^2>1$ for $l=m,\dots,k$.  Hence
\begin{equation}\label{eq:k_odd_first3}
  \beta^2(j+1-2k)+1<0,
\end{equation}
as required. Now take $j=m,\dots,\frac{k-1}{2}$.  By assumption, $l\beta^2>1$ for $l\geq m$, and $m\leq\frac{k-1}{2}$. We therefore have
\begin{equation}
  j\beta^2(k-j)>1>\frac{1}{2}
\end{equation}
hence
\begin{equation}
  2j\beta^2(j-k)+1<0.
\end{equation}

If $k$ is even, but still $2<m\leq\frac{k}{2}$ then entirely similarly to the odd $k$ case we find
\begin{align}
  \mom_n(k,\beta)\sim2^{(k^2\beta^2-k+1)n}&\Bigg[\sum_{j=1}^{m-1}\binom{k}{j}\rho(j,\beta)\tau(k-j,\beta)2^{j\beta^2(j-1)-j}\frac{1-2^{j(\beta^2(j+1-2k)+1)n}}{2^{j(\beta^2(2k-j-1)-1)}-1}\nonumber\\
  &\quad+\sum_{j=m}^{\frac{k-2}{2}}\binom{k}{j}\tau(j,\beta)\tau(k-j,\beta)\frac{1-2^{(2j\beta^2(j-k)+1)n}}{2(2^{2j\beta^2(k-j)-1}-1)}\nonumber\\
  &\quad+\binom{k}{\frac{k}{2}}\tau(\tfrac{k}{2},\beta)^2\frac{1-2^{(1-\frac{k^2\beta^2}{2})n}}{2^2(2^{\frac{k^2\beta^2}{2}-1}-1)}+1\Bigg],\label{eq:k_even_first1}
\end{align}
thus, the same arguments hold for the first two sums of \eqref{eq:k_even_first1} as for the odd $k$ case.  We are done provided additionally that the contribution from $2^{(1-\frac{k^2\beta^2}{2})n}$ is subleading. As $k\beta^2>1$ and $k>2$, we therefore have $k^2\beta^2>2$, and so
\begin{equation}
  1-\frac{k^2\beta^2}{2}<0.
\end{equation}

Moving to the case where $\frac{k-1}{2}<m\leq k$ and $k$ odd, and applying \eqref{eq:induction_hyp} to \eqref{eq:k_odd_second} we find
\begin{align}
  \mom_n(k,\beta)&\sim 
  2^{(k^2\beta^2-k+1)n}\Bigg[\sum_{j=1}^{k-m}\binom{k}{j}\rho(j,\beta)\tau(k-j,\beta)2^{j(\beta^2(j-1)-1)}\frac{1-2^{j(\beta^2(j+1-2k)+1)n}}{2^{j(\beta^2(2k-j-1)-1)}-1} +1\Bigg]\nonumber\\
  &\quad+2^{k^2\beta^2-k+1-k\beta^2}\sum_{j=k-m+1}^{\frac{k-1}{2}}\binom{k}{j}\rho(j,\beta)\rho(k-j,\beta)2^{2j\beta^2(j-k)}\frac{2^{(k^2\beta^2-k+1)n}-2^{k\beta^2n}}{2^{k^2\beta^2-k+1-k\beta^2}-1}.\label{eq:k_odd_second1}
\end{align}
Since $k\beta^2>1$ by assumption, we have that the terms in the second sum of \eqref{eq:k_odd_second1} do grow asymptotically like $2^{(k^2\beta^2-k+1)n}$.  To confirm that this is also true for the terms in the first sum, we check that $j(\beta^2(j+1-2k)+1<0$ for $j=1,\dots,k-m$ and $\frac{k-1}{2}<m\leq k$.  This is true by the same arguments as above (see \eqref{eq:k_odd_first2} and \eqref{eq:k_odd_first3}). 

To conclude we consider the case of even $k$ and $\frac{k}{2}<m\leq k$, we find here that
\begin{align}
  \mom_n(k,\beta)
  &\sim 2^{(k^2\beta^2-k+1)n}\Bigg[\sum_{j=1}^{k-m}\binom{k}{j}\rho(j,\beta)\tau(k-j,\beta)2^{j(\beta^2(j-1)-1)}\frac{1-2^{j(\beta^2(j+1-2k)+1)n}}{2^{j(\beta^2(2k-j-1)-1)}-1} +1\Bigg]\nonumber\\
  &\quad+2^{k^2\beta^2-k+1-k\beta^2}\sum_{j=k-m+1}^{\frac{k-2}{2}}\binom{k}{j}\rho(j,\beta)\rho(k-j,\beta)2^{2j\beta^2(j-k)}\frac{2^{(k^2\beta^2-k+1)n}-2^{k\beta^2n}}{2^{k^2\beta^2-k+1-k\beta^2}-1}\nonumber\\
  &\quad+2^{\frac{k^2\beta^2}{2}-k-k\beta^2}\binom{k}{\frac{k}{2}}\rho(\tfrac{k}{2},\beta)^2\frac{2^{(k^2\beta^2-k+1)n}-2^{k\beta^2n}}{2^{k^2\beta^2-k+1-k\beta^2}-1},\label{eq:k_even_second1}
\end{align}
thus we employ the arguments of \eqref{eq:k_odd_second1}.  This concludes the proof for $\abs{\beta}\in\left(\frac{1}{\sqrt{m}},\frac{1}{\sqrt{m-1}}\right)$, where $m\in\{3,\dots,k\}$.

\subsubsection{Range: $\frac{1}{\sqrt{2}}<|\beta|$}\label{sec:beta_2}

In this range, $l\beta^2>1$ for all $l=2,\dots,k$.  Since $\mom_n(1,\beta)=2^{\beta^2n}$, we replace all occurrences of $\mom_{n-\lambda-1}(l,\beta)$ in \eqref{eq:inductive_decomp1} by
\begin{equation}
  \tau(j,\beta)2^{(j^2\beta^2-j+1)(n-\lambda-1)}
\end{equation}
for $j=1,\dots,k-1$ using \eqref{eq:induction_hyp} (where recall we define $\tau(1,\beta)\equiv 1)$).  Thus,

\begin{align}
  \mom_n(k,\beta)
  &\sim 2^{(k^2\beta^2-k+1)n}\Bigg[\frac{1}{2^2}\sum_{j=1}^{k-1}\binom{k}{j}\tau(j,\beta)\tau(k-j,\beta)2^{(2j\beta^2(j-k)+1)n}\sum_{\lambda=0}^{n-1}2^{(2j\beta^2(k-j)-1)}\lambda+1\Bigg]\nonumber\\
  &= 2^{(k^2\beta^2-k+1)n}\Bigg[\frac{1}{2^2}\sum_{j=1}^{k-1}\binom{k}{j}\tau(j,\beta)\tau(k-j,\beta)\frac{1-2^{(2j\beta^2(j-k)+1)n}}{2^{2j\beta^2(k-j)-1}-1}+1\Bigg].
\end{align}

Thus, in order to establish that $\mom_n(k,\beta)$ grows as $2^{(k^2\beta^2-k+1)n}$ for this range of $\beta$, we confirm that the contribution from $2^{(2j\beta^2(j-k)+1)n}$ is subleading for $j=1,\dots,k-1$.  This is true since $2j\beta^2>1$ by assumption and $k-j\geq1$, so
\begin{equation}
  2j\beta^2(k-j)>1.
\end{equation}

\subsubsection{Range: $|\beta|=\frac{1}{\sqrt{m}}$}\label{sec:beta_3}
Assume that $\abs{\beta}=\frac{1}{\sqrt{m}}$ for some $m\in\{2,\dots,k-1\}$ (the cases $m=1,k$ were dealt with in sections~\ref{sec:beta_2} and~\ref{sec:second_regime} respectively).  We revisit the techniques used in section~\ref{sec:beta_1}.  Beginning with the odd $k$ case, we separate \eqref{eq:k_odd} around the $m$th term to find

\begin{align}
  \mom_n(k,\beta)
  &=2^{k^2\beta^2-k+1}\sum_{\lambda=0}^{n-1}2^{(k^2\beta^2-k+1)\lambda}\nonumber\\
  &\qquad\Bigg[\sum_{j=1}^{m-1}\binom{k}{j}2^{2j\beta^2(j-k)}\mom_{n-\lambda-1}(j,\beta)\mom_{n-\lambda-1}(k-j,\beta)\nonumber\\
    &\qquad +\binom{k}{m}2^{2\beta^2m(m-k)}\mom_{n-\lambda-1}(m,\beta)\mom_{n-\lambda-1}(k-m,\beta)\nonumber\\
    &\qquad+\sum_{j=m+1}^{\frac{k-1}{2}}\binom{k}{j}2^{2j\beta^2(j-k)}\mom_{n-\lambda-1}(j,\beta)\mom_{n-\lambda-1}(k-j,\beta)\Bigg]\nonumber\\
  &\qquad+2^{(k^2\beta^2-k+1)n}.
\end{align}

Since the sums over $j=1,\dots,m-1$ and $j=m+1,\dots,\frac{k-1}{2}$ can be handled using the arguments of section~\ref{sec:beta_1}, we only need to determine how the middle term (i.e. $j=m$) grows asymptotically. Using \eqref{eq:induction_hyp} we examine
\begin{align}
  2^{k^2\beta^2-k+1}&\binom{k}{m}2^{2\beta^2m(m-k)}\sum_{\lambda=0}^{n-1}2^{(k^2\beta^2-k+1)\lambda}\mom_{n-\lambda-1}(m,\beta)\mom_{n-\lambda-1}(k-m,\beta)\nonumber\\
  &\sim\binom{k}{m}\sigma(m,\tfrac{1}{\sqrt{m}})\tau(k-m,\tfrac{1}{\sqrt{m}})2^{k^2\beta^2-k+1+2(m-k)}\nonumber\\
  &\qquad\sum_{\lambda=0}^{n-1}(n-\lambda-1)2^{(k^2\beta^2-k+1)\lambda}2^{(1+(k-m)^2\beta^2-(k-m)+1)(n-\lambda-1)}\\
  &=\frac{1}{2}\binom{k}{m}\sigma(m,\tfrac{1}{\sqrt{m}})\tau(k-m,\tfrac{1}{\sqrt{m}})2^{(k^2\beta^2-k+1+2(m-k)+1)n}\sum_{\lambda=0}^{n-1}(n-\lambda-1)2^{(2(k-m)-1)\lambda}.\label{eq:middle_j_1}
\end{align}

Computing the sum over $\lambda$, we have
\begin{align}
  \sum_{\lambda=0}^{n-1}(n-\lambda-1)&2^{2(m-k)\lambda}
  =\frac{2^{(2(k-m)-1)n}-1+n(1-2^{2(k-m)-1})}{(1-2^{2(k-m)-1})^2}.\label{eq:middle_j_2}
\end{align}

Combining \eqref{eq:middle_j_1} and \eqref{eq:middle_j_2} gives
\begin{align}
  \frac{1}{2}\binom{k}{m}&\sigma(m,\tfrac{1}{\sqrt{m}})\tau(k-m,\tfrac{1}{\sqrt{m}})2^{(k^2\beta^2-k+1+2(m-k)+1)n}\sum_{\lambda=0}^{n-1}(n-\lambda-1)2^{(2(k-m)-1)\lambda}\nonumber\\
  &=\frac{1}{2}\binom{k}{m}\sigma(m,\tfrac{1}{\sqrt{m}})\tau(k-m,\tfrac{1}{\sqrt{m}})\nonumber\\
  &\qquad\times\frac{2^{(k^2\beta^2-k+1)n}-2^{(k^2\beta^2-k+1+2(m-k)+1)n}+n2^{(k^2\beta^2-k+1+2(m-k)+1)n}(1-2^{2(k-m)-1})}{(1-2^{2(k-m)-1})^2}.
\end{align}

Thus, the result follows once it is established that $2(m-k)+1$ is negative.  By assumption $m\in\{2,\dots,k-1\}$, so $2(k-m)>1$ and thus we conclude.  The case for even $k$ follows from precisely the same reasoning, except in the case where $m=\frac{k}{2}$.

Assume $k\geq 4$ is even\footnote{Recall that the case $k=2$ has already been calculated in section~\ref{sec:structure}.} and $m=\frac{k}{2}$, then by \eqref{eq:k_even} we have
\begin{align}
  \mom_n(k,\beta)&=2^{k^2\beta^2-k+1}\sum_{\lambda=0}^{n-1}2^{(k^2\beta^2-k+1)\lambda}\sum_{j=1}^{\frac{k-2}{2}}\binom{k}{j}2^{2j\beta^2(j-k)}\nonumber\\
  &\qquad\quad\times \mom_{n-\lambda-1}(j,\beta)\mom_{n-\lambda-1}(k-j,\beta)\nonumber\\
  &\quad +2^{\frac{k^2\beta^2}{2}-k}\binom{k}{\frac{k}{2}}\sum_{\lambda=0}^{n-1}2^{(k^2\beta^2-k+1)\lambda}\left(\mom_{n-\lambda-1}(\tfrac{k}{2},\beta)\right)^2+2^{(k^2\beta^2-k+1)n}.
\end{align}

As above, the sum has already been handled by the argument in section~\ref{sec:beta_1}, thus we only consider the penultimate term. Since $k\geq 4$, when we apply \eqref{eq:induction_hyp}, we may replace $\mom_{n-\lambda-1}(\frac{k}{2},\beta)$ with $\sigma(k,\beta)(n-\lambda-1)2^{n-\lambda-1}$ (this is not true if $k=2$):
\begin{align}
  2^{\frac{k^2\beta^2}{2}-k}\binom{k}{\frac{k}{2}}\sum_{\lambda=0}^{n-1}&2^{(k^2\beta^2-k+1)\lambda}\left(\mom_{n-\lambda-1}\left(\tfrac{k}{2},\beta\right)\right)^2\nonumber\\
  &\sim\binom{k}{\frac{k}{2}}\sigma(k,\beta)^22^{2(n-1)}\sum_{\lambda=0}^{n-1}(n-\lambda-1)^22^{(k-1)\lambda}\\
  &=\binom{k}{\frac{k}{2}}\sigma(k,\beta)^2\frac{(2^{(k+1)n}-2^{2n})(2^{k}+2)-2(n2^{n}(2^{k-1}-1))^2-n2^{2n}(2^{k}-2)}{(2^{k}-2)^3}.
\end{align}

By assumption, $\frac{k}{2}\beta^2=1$, so $k^2\beta^2-k+1$ simplifies to $k+1$. Thus, the result follows if $2^{2n}$ is subleading to $2^{(k+1)n}$.  This follows since $k\geq 4$.  

\subsection{Proof of Corollary~\ref{cor:branching_mom}}

Assume $k, \beta\in\mathbb{N}$. When $k=1, 2$, the result follows from the computation in section~\ref{sec:structure}, see \eqref{eq:k=1a}, \eqref{eq:k=1b} and \eqref{eq:k2_sum}.  There it is shown that
\begin{align}
  \mom_n(1,\beta)&=2^{\beta^2n},\\
  \mom_n(2,\beta)&=\frac{2^{2\beta^2}-1}{2^{2\beta^2}-2}2^{(4\beta^2-1)n}-\frac{1}{2^{2\beta^2}-2}2^{(2\beta^2)n},
\end{align}
thus both are polynomials in $2^n$ of degree $k^2\beta^2-k+1$. Proceeding inductively, assume that
\begin{equation}
  \mom_n(j,\beta)=\sum_{r_j=0}^{j^2\beta^2-j+1}\alpha^j_{r_j}2^{nr_j}
\end{equation}
for $j<k$, and that $\alpha^j_{r_j}$ are the appropriate coefficients for the polynomial $\mom_n(j,\beta)$. Then, using \eqref{eq:inductive_decomp1},
\begin{align}
  &\mom_n(k,\beta)\nonumber\\
  &\qquad=2^{k^2\beta^2-k}\sum_{j=1}^{k-1}\binom{k}{j}2^{2j\beta^2(j-k)}\nonumber\\
  &\qquad\qquad\quad\times\sum_{\lambda=0}^{n-1}2^{(k^2\beta^2-k+1)\lambda} \mom_{n-\lambda-1}(j,\beta)\mom_{n-\lambda-1}(k-j,\beta)+2^{(k^2\beta^2-k+1)n}\\
  &\qquad=2^{k^2\beta^2-k}\sum_{j=1}^{k-1}\binom{k}{j}2^{2j\beta^2(j-k)}\sum_{r_j=0}^{j^2\beta^2-j+1}\sum_{r_{k-j}=0}^{(k-j)^2\beta^2-(k-j)+1}\alpha^j_{r_j}\alpha^{k-j}_{r_{k-j}}2^{(r_j+r_{k-j})(n-1)}\nonumber\\
  &\qquad\qquad\quad\times \sum_{\lambda=0}^{n-1}2^{(k^2\beta^2-k+1-r_j-r_{k-j})\lambda}+2^{(k^2\beta^2-k+1)n}\\
  &\qquad=2^{k^2\beta^2-k}\sum_{j=1}^{k-1}\binom{k}{j}2^{2j\beta^2(j-k)}\sum_{r_j=0}^{j^2\beta^2-j+1}\sum_{r_{k-j}=0}^{(k-j)^2\beta^2-(k-j)+1}\alpha^j_{r_j}\alpha^{k-j}_{r_{k-j}}2^{(r_j+r_{k-j})(n-1)}\nonumber\\
  &\qquad\qquad\quad\times\frac{2^{(k^2\beta^2-k+1-(r_j-r_{k-j}))n}-1}{2^{k^2\beta^2-k+1-(r_j+r_{k-j})}-1}+2^{(k^2\beta^2-k+1)n}\\
  &\qquad=2^{k^2\beta^2-k}\sum_{j=1}^{k-1}\sum_{r_j=0}^{j^2\beta^2-j+1}\sum_{r_{k-j}=0}^{(k-j)^2\beta^2-(k-j)+1}\frac{\binom{k}{j}2^{2j\beta^2(j-k)}\alpha^j_{r_j}\alpha^{k-j}_{r_{k-j}}}{2^{k^2\beta^2-k+1}-2^{r_j+r_{k-j}}}(2^{(k^2\beta^2-k+1)n}-2^{(r_j+r_{k-j})n})\nonumber\\
  &\qquad\qquad\quad+2^{(k^2\beta^2-k+1)n}.
\end{align}
Thus, $\mom_n(k,\beta)$ is a sum of polynomials in $2^n$.  Since we determined in section~\ref{sec:third_regime} that when $k\beta^2>1$, $\mom_n(k,\beta)$ grows like $2^{(k^2\beta^2-k+1)n}$, we hence have shown that $\mom_n(k,\beta)$ is a polynomial in $2^n$ of this degree.  This completes the proof of corollary~\ref{cor:branching_mom}.

\section{Acknowledgements}\label{sec:acknowledgements}

ECB is grateful to the Heilbronn Institute for Mathematical Research for support.  JPK is pleased to acknowledge support from ERC Advanced Grant 740900 (LogCorRM) and a Royal Society Wolfson Research Merit Award.  We would also like to thank the anonymous referees for their careful reading of the manuscript, and for their helpful questions and comments. 

\appendix
\section{Explicit examples for small $k$}\label{sec:appendix}

We completely determined $\mom_n(1,\beta)$ and $\mom_n(2,\beta)$ in section~\ref{sec:structure}, see ~\eqref{eq:k=1b}, \eqref{eq:mom2_1}, \eqref{eq:mom2_2}, \eqref{eq:mom2_3}. These are valid for any non-zero, real $\beta$ (and $\mom_n(k,0)=1$ for any $k\in\mathbb{N}$).  Hence, we have
\begin{equation}
  \mom_n(1,\beta)=2^{\beta^2n}
\end{equation}
and as $n\rightarrow\infty$
\begin{equation}
  \mom_n(2,\beta)\sim
  \begin{cases}
    \frac{1}{2(1-2^{2\beta^2-1})}2^{2\beta^2n},&\text{if }2\beta^2<1,\\
    \frac{n}{2}2^n,&\text{if }2\beta^2=1,\\
    \frac{2^{2\beta^2}-1}{2(2^{2\beta^2-1}-1)}2^{(4\beta^2-1)n},&\text{if }2\beta^2>1.
  \end{cases}
\end{equation}

Following the method outlined in section~\ref{sec:structure}, we calculate for $k=3$:
\begin{align}
  \mom_n(3,\beta)&=\frac{1}{2^{3n}}\sum_{l_1,l_2,l_3=0}^{2^n}\mathbb{E}\left[e^{2\beta(X_n(l_1)+X_n(l_2)+X_n(l_3))}\right]\\
    &=\frac{2}{2^{3n}}\binom{3}{1}\sum_{\lambda=0}^{n-1}2^{(9\beta^2+1)\lambda}2^{5\beta^2}2^{3(n-\lambda-1)}2^{\beta^2(n-\lambda-1)}\mom_{n-\lambda-1}(2,\beta)+2^{(9\beta^2-2)n}\\
  &=\binom{3}{1}2^{4\beta^2-2}2^{\beta^2n}\sum_{\lambda=0}^{n-1}2^{(8\beta^2-2)\lambda}\left(2^{2\beta^2(n-\lambda-1)}\frac{2^{(2\beta^2-1)(n-\lambda-1)}-1}{2(2^{2\beta^2-1}-1)}+2^{(4\beta^2-1)(n-\lambda-1)}\right)\nonumber\\
  &\qquad+2^{(9\beta^2-2)n}.\label{eq:mom3_1}
\end{align}
The second equality follows by splitting the paths before and after the last common level $\lambda\coloneqq\lcl(l_1,l_2,l_3)$.  Note that at the splitting point, there are $2^\lambda$ choices for $\lca(l_1,l_2,l_3)$ given $\lambda$, and hence $\binom{3}{1}$ choices for which paths are paired, and which is the single path. The third equality follows by substituting in \eqref{eq:k2_sum}. Hence, calculating the sum over $\lambda$, we find

\begin{align}
  \mom_n(3,\beta)
  &=\binom{3}{1}\Bigg[\frac{2^{(9\beta^2-2)n}-2^{(5\beta^2-1)n}}{2(2^{4\beta^2-1}-1)}+\frac{2^{(9\beta^2-2)n}-2^{(5\beta^2-1)n}}{2(2^{2\beta^2}-2)(2^{4\beta^2-1}-1)}\nonumber\\
    &\qquad\qquad-\frac{2^{2\beta^2}(2^{(9\beta^2-2)n}-2^{3\beta^2n})}{2^{2}(2^{2\beta^2}-2)(2^{6\beta^2-2}-1)}\Bigg]+2^{(9\beta^2-2)n}\\
  &=\binom{3}{1}\Bigg[\frac{2^{6\beta^2}-2}{(2^{4\beta^2}-2)(2^{6\beta^2}-2^2)}2^{(9\beta^2-2)n} -\frac{2^{2\beta^2}-1}{(2^{4\beta^2}-2)(2^{2\beta^2}-2)}2^{(5\beta^2-1)n}\nonumber\\
  &\qquad\qquad +\frac{2^{2\beta^2}}{(2^{6\beta^2}-2^2)(2^{2\beta^2}-2)}2^{3\beta^2n}\Bigg]+2^{(9\beta^2-2)n}.
\end{align}

Thus, if $3\beta^2>1$ then
\begin{equation}
  \mom_n(3,\beta)\sim \left(1+\frac{3(2^{6\beta^2}-2)}{(2^{4\beta^2}-2)(2^{6\beta^2}-2^2)}\right)2^{(9\beta^2-2)n}.
\end{equation}
If instead $3\beta^2<1$ then
\begin{equation}
  \mom_n(3,\beta)\sim \frac{3\cdot2^{2\beta^2}}{(2^{6\beta^2}-2^2)(2^{2\beta^2}-2)}2^{3\beta^2n}.
\end{equation}
Finally, if $3\beta^2=1$ then using \eqref{eq:mom3_1} we find
\begin{align}
  \mom_n(3,\tfrac{1}{\sqrt{3}})&=3\cdot 2^{\frac{1}{3}(n-2)}\Bigg[\frac{2^{\frac{2}{3}(n-1)}}{2^{\frac{2}{3}}-2}\sum_{\lambda=0}^{n-1}(2^{-\frac{1}{3}(n-\lambda-1)}-1)+2^{\frac{1}{3}(n-1)}\sum_{\lambda=0}^{n-1}2^{\frac{1}{3}\lambda}\Bigg]+2^n\\
  &=\frac{3}{2^{\frac{7}{3}}-2^2}n2^n-\frac{3\cdot 2^{\frac{1}{3}}}{(2^{\frac{7}{3}}-2^2)(2^{\frac{1}{3}}-1)}(2^n-2^{\frac{2}{3}n})+\frac{3}{2^{\frac{4}{3}}-2}(2^n-2^{\frac{2}{3}n})+2^n.
\end{align}
Thus, as $n\rightarrow\infty$ 
\begin{equation}
  \mom_n(3,\tfrac{1}{\sqrt{3}})\sim \frac{3}{2^{\frac{7}{3}}-2^2}n2^n.
\end{equation}

Hence, overall as $n\rightarrow\infty$,
\begin{equation}
  \mom_n(3,\beta)\sim
  \begin{cases}
    \frac{3\cdot2^{2\beta^2}}{(2^2-2^{6\beta^2})(2-2^{2\beta^2})}2^{3\beta^2n},&\text{if }3\beta^2<1,\\
    \frac{3}{2^{\frac{7}{3}}-2^2}n2^n,&\text{if }3\beta^2=1,\\
    \left(1+\frac{3(2^{6\beta^2}-2)}{(2^{4\beta^2}-2)(2^{6\beta^2}-2^2)}\right)2^{(9\beta^2-2)n},&\text{if }3\beta^2>1.
  \end{cases}
\end{equation}

Using the inductive method described above, one additionally finds that
\begin{align}
  \mom_n(4,\beta)&=2^{\left(16 \beta^2-3\right) n}+\binom{4}{1}\frac{2^{\left(16 \beta^2-3\right) n}-2^{\left(10 \beta^2-2\right) n}}{2^{6 \beta^2}-2}\nonumber\\
  &\quad+\binom{4}{1}\binom{3}{1} \Bigg[\frac{2^{\left(16 \beta^2-3\right) n}-2^{\left(10 \beta^2-2\right) n}}{\left(2^{4 \beta^2}-2\right) \left(2^{6\beta^2}-4\right)}+\frac{2^{8 \beta^2} \left(2^{\left(16 \beta^2-3\right) n}-2^{4 \beta^2n}\right)}{\left(2^{2 \beta^2}-2\right) \left(2^{6 \beta^2}-4\right) \left(2^{12\beta^2}-8\right)}\nonumber\\
  &\qquad\qquad\qquad\qquad-\frac{\left(2^{6 \beta^2}-2^{4 \beta^2}\right) \left(2^{\left(16\beta^2-3\right) n}-2^{\left(6 \beta^2-1\right) n}\right)}{\left(2^{2 \beta^2}-2\right)\left(2^{4 \beta^2}-2\right) \left(2^{10 \beta^2}-4\right)}\Bigg]\nonumber\\
  &\quad+\binom{4}{2}2^{8 \beta^2-4}\Bigg[\frac{2^{\left(16 \beta^2-3\right) n}-2^{4 \beta^2 n}}{\left(2^{2 \beta^2}-2\right)^2\left(2^{16 \beta^2-3}-2^{4 \beta^2}\right)}-\frac{2^{2-6 \beta^2} \left(2^{\left(16\beta^2-3\right) n}-2^{\left(6 \beta^2-1\right) n}\right)}{\left(2^{2 \beta^2}-2\right)^2\left(2^{10 \beta^2-2}-1\right)}\nonumber\\
  &\qquad\qquad\qquad\qquad-\frac{2^{\left(16 \beta^2-3\right) n}-2^{\left(6\beta^2-1\right) n}}{\left(2^{2 \beta^2}-2\right) \left(2^{16 \beta^2-3}-2^{6\beta^2-1}\right)}+\frac{2^{2-8 \beta^2} \left(2^{\left(16 \beta^2-3\right) n}-2^{\left(8\beta^2-2\right) n}\right)}{\left(2^{2 \beta^2}-2\right)^2 \left(2^{8\beta^2-1}-1\right)}\nonumber\\
  &\qquad\qquad\qquad\qquad+\frac{2^{\left(16 \beta^2-3\right) n}-2^{\left(8 \beta^2-2\right)n}}{\left(2^{2 \beta^2}-2\right) \left(2^{16 \beta^2-3}-2^{8\beta^2-2}\right)}+\frac{2^{\left(16 \beta^2-3\right) n}-2^{\left(8 \beta^2-2\right) n}}{2^{16\beta^2-3}-2^{8 \beta^2-2}}\Bigg].
\end{align}

Thus, if $4\beta^2<1$ then
\begin{equation}
  \frac{\mom_n(4,\beta)}{2^{4\beta^2n}}\sim \frac{3\cdot2^{8 \beta^2+2}}{\left(2-2^{2 \beta^2}\right) \left(4-2^{6\beta^2}\right) \left(8-2^{12 \beta^2}\right)}+\frac{3\cdot 2^{8 \beta^2-3}}{\left(2-2^{2 \beta^2}\right)^2 \left(2^{4 \beta^2}-2^{16\beta^2-3}\right)}
\end{equation}
as $n\rightarrow\infty$. Instead if $4\beta^2>1$ then
\begin{align}
  \frac{\mom_n(4,\beta)}{2^{(16\beta^2-3)n}}&\sim 1+\frac{4}{2^{6 \beta^2}-2}\nonumber\\
  &\quad+12 \Bigg[\frac{1}{\left(2^{4 \beta^2}-2\right)\left(2^{6 \beta^2}-4\right)}-\frac{2^{6 \beta^2}-2^{4 \beta^2}}{\left(2^{2 \beta^2}-2\right) \left(2^{4\beta^2}-2\right) \left(2^{10 \beta^2}-4\right)}\nonumber\\
  &\qquad\qquad\qquad\qquad+\frac{2^{8 \beta^2}}{\left(2^{2 \beta^2}-2\right) \left(2^{6\beta^2}-4\right) \left(2^{12 \beta^2}-8\right)}\Bigg]\nonumber\\
  &\quad+3\cdot2^{8 \beta^2-3} \Bigg[\frac{1}{\left(2^{2 \beta^2}-2\right)^2 \left(2^{16 \beta^2-3}-2^{4 \beta^2}\right)}-\frac{2^{2-6 \beta^2}}{\left(2^{2 \beta^2}-2\right)^2 \left(2^{10 \beta^2-2}-1\right)}\nonumber\\
    &\qquad\qquad\qquad\qquad-\frac{1}{\left(2^{2\beta^2}-2\right) \left(2^{16 \beta^2-3}-2^{6 \beta^2-1}\right)}+\frac{1}{\left(2^{2 \beta^2}-2\right) \left(2^{16 \beta^2-3}-2^{8 \beta^2-2}\right)}\nonumber\\
    &\qquad\qquad\qquad\qquad+\frac{1}{2^{16 \beta^2-3}-2^{8\beta^2-2}}+\frac{2^{2-8 \beta^2}}{\left(2^{2 \beta^2}-2\right)^2 \left(2^{8 \beta^2-1}-1\right)}\Bigg].
\end{align}

For $k=5$ we have
\begingroup
\allowdisplaybreaks
\begin{align}
  &\mom_n(5,\beta)\nonumber\\
  &\quad=\binom{5}{1}2^{16\beta^2-4}2^{\beta^2n}\nonumber\\
  &\quad\quad\Bigg[\frac{2^{(16\beta^2-3)(n-1)}\left(2^{(8\beta^2-1)n}-1\right)}{2^{8\beta^2-1}-1}+\frac{\binom{4}{1}2^{(16\beta^2-3)(n-1)}\left(2^{(8\beta^2-1)n}-1\right)}{(2^{6\beta^2}-2)(2^{8\beta^2-1}-1)}\nonumber\\
    &\quad\qquad-\frac{\binom{4}{1}2^{(10\beta^2-2)(n-1)}\left(2^{(14\beta^2-2)n}-1\right)}{(2^{6\beta^2}-2)(2^{14\beta^2-2}-1)}+\frac{\binom{4}{1}\binom{3}{1}2^{(16\beta^2-3)(n-1)}\left(2^{(8\beta^2-1)n}-1\right)}{(2^{4\beta^2}-2)(2^{6\beta^2}-4)(2^{8\beta^2-1}-1)}\nonumber\\
    &\quad\qquad-\frac{\binom{4}{1}\binom{3}{1}2^{(10\beta^2-2)(n-1)}\left(2^{(14\beta^2-2)n}-1\right)}{(2^{4\beta^2}-2)(2^{6\beta^2}-4)(2^{14\beta^2-2}-1)}-\frac{\binom{4}{1}\binom{3}{1}(2^{6\beta^2}-2^{4\beta^2})2^{(16\beta^2-3)(n-1)}\left(2^{(8\beta^2-1)n}-1\right)}{(2^{4\beta^2}-2)(2^{2\beta^2}-2)(2^{10\beta^2}-4)(2^{8\beta^2-1}-1)}\nonumber\\
    &\quad\qquad+\frac{\binom{4}{1}\binom{3}{1}(2^{6\beta^2}-2^{4\beta^2})2^{(6\beta^2-1)(n-1)}\left(2^{(18\beta^2-3)n}-1\right)}{(2^{4\beta^2}-2)(2^{2\beta^2}-2)(2^{10\beta^2}-4)(2^{18\beta^2-3}-1)}+\frac{\binom{4}{1}\binom{3}{1}2^{8\beta^2}2^{(16\beta^2-3)(n-1)}\left(2^{(8\beta^2-1)n}-1\right)}{(2^{2\beta^2}-2)(2^{6\beta^2}-4)(2^{12\beta^2}-8)(2^{8\beta^2-1}-1)}\nonumber\\
    &\quad\qquad-\frac{\binom{4}{1}\binom{3}{1}2^{8\beta^2}2^{4\beta^2(n-1)}\left(2^{(20\beta^2-4)n}-1\right)}{(2^{2\beta^2}-2)(2^{6\beta^2}-4)(2^{12\beta^2}-8)(2^{20\beta^2-4}-1)}+\frac{\binom{4}{2}2^{8\beta^2}2^{(16\beta^2-3)(n-1)}\left(2^{(8\beta^2-1)n}-1\right)}{2^4(2^{2\beta^2}-2)^2(2^{16\beta^2-3}-2^{4\beta^2})(2^{8\beta^2-1}-1)}\nonumber\\
    &\quad\qquad-\frac{\binom{4}{2}2^{8\beta^2}2^{4\beta^2(n-1)}\left(2^{(20\beta^2-4)n}-1\right)}{2^4(2^{2\beta^2}-2)^2(2^{16\beta^2-3}-2^{4\beta^2})(2^{20\beta^2-4}-1)}+\frac{\binom{4}{2}2^{(16\beta^2-3)(n-1)}\left(2^{(8\beta^2-1)n}-1\right)}{2^2(2^{2\beta^2}-2)^2(2^{8\beta^2-1}-1)^2}\nonumber\\
    &\quad\qquad-\frac{\binom{4}{2}2^{(8\beta^2-2)(n-1)}\left(2^{(16\beta^2-2)n}-1\right)}{2^2(2^{2\beta^2}-2)^2(2^{8\beta^2-1}-1)(2^{16\beta^2-2}-1)}+\frac{\binom{4}{2}2^{8\beta^2}2^{(16\beta^2-3)(n-1)}\left(2^{(8\beta^2-1)n}-1\right)}{2^{8\beta^2+2}(2^{8\beta^2-1}-1)^2}\nonumber\\
    &\quad-\frac{\binom{4}{2}2^{8\beta^2}2^{(8\beta^2-2)(n-1)}\left(2^{(16\beta^2-2)n}-1\right)}{2^{8\beta^2+2}(2^{8\beta^2-1}-1)(2^{16\beta^2-2}-1)}+\frac{\binom{4}{2}2^{8\beta^2}2^{(16\beta^2-3)(n-1)}\left(2^{(8\beta^2-1)n}-1\right)}{2^{8\beta^2+2}(2^{2\beta^2}-2)(2^{8\beta^2-1}-1)^2}\nonumber\\
    &\quad\qquad-\frac{\binom{4}{2}2^{8\beta^2}2^{(8\beta^2-2)(n-1)}\left(2^{(16\beta^2-2)n}-1\right)}{2^{8\beta^2+2}(2^{2\beta^2}-2)(2^{8\beta^2-1}-1)(2^{16\beta^2-2}-1)}-\frac{\binom{4}{2}2^{2\beta^2}2^{(16\beta^2-3)(n-1)}\left(2^{(8\beta^2-1)n}-1\right)}{2^{2}(2^{2\beta^2}-2)^2(2^{10\beta^2-2}-1)(2^{8\beta^2-1}-1)}\nonumber\\
    &\quad\qquad+\frac{\binom{4}{2}2^{2\beta^2}2^{(6\beta^2-1)(n-1)}\left(2^{(18\beta^2-3)n}-1\right)}{2^{2}(2^{2\beta^2}-2)^2(2^{10\beta^2-2}-1)(2^{18\beta^2-3}-1)}-\frac{\binom{4}{2}2^{2\beta^2}2^{(16\beta^2-3)(n-1)}\left(2^{(8\beta^2-1)n}-1\right)}{2^{3}(2^{2\beta^2}-2)(2^{10\beta^2-2}-1)(2^{8\beta^2-1}-1)}\nonumber\\
    &\quad\qquad+\frac{\binom{4}{2}2^{2\beta^2}2^{(6\beta^2-1)(n-1)}\left(2^{(18\beta^2-3)n}-1\right)}{2^{3}(2^{2\beta^2}-2)(2^{10\beta^2-2}-1)(2^{18\beta^2-3}-1)}\Bigg]\nonumber\\
  &\quad\quad+\binom{5}{2}2^{13\beta^2-4}\nonumber\\
  &\quad\Bigg[\frac{\binom{3}{1}(2^{2\beta^2}-1)(2^{6\beta^2}-2)2^{(13\beta^2-3)(n-1)}(2^{(12\beta^2-1)n}-1)}{(2^{2\beta^2}-2)(2^{6\beta^2}-4)(2^{4\beta^2}-2)(2^{12\beta^2-1}-1)}-\frac{\binom{3}{1}(2^{2\beta^2}-1)^22^{(9\beta^2-2)(n-1)}(2^{(16\beta^2-2)n}-1)}{(2^{2\beta^2}-2)^2(2^{4\beta^2}-2)(2^{16\beta^2-2}-1)}\nonumber\\
    &\quad\qquad+\frac{\binom{3}{1}(2^{4\beta^2}-2^{2\beta^2})2^{(7\beta^2-1)(n-1)}(2^{(18\beta^2-3)n}-1)}{(2^{2\beta^2}-2)^2(2^{6\beta^2}-4)(2^{18\beta^2-3}-1)}+\frac{(2^{2\beta^2}-1)2^{(13\beta^2-3)(n-1)}(2^{(12\beta^2-1)n}-1)}{(2^{2\beta^2}-2)(2^{12\beta^2-1}-1)}\nonumber\\
    &\quad\qquad-\frac{\binom{3}{1}(2^{6\beta^2}-2)2^{(11\beta^2-2)(n-1)}(2^{(14\beta^2-2)n}-1)}{(2^{2\beta^2}-2)(2^{6\beta^2}-4)(2^{4\beta^2}-2)(2^{14\beta^2-2}-1)}+\frac{\binom{3}{1}(2^{2\beta^2}-1)2^{(7\beta^2-1)(n-1)}(2^{(18\beta^2-3)n}-1)}{(2^{2\beta^2}-2)^2(2^{4\beta^2}-2)(2^{18\beta^2-3}-1)}\nonumber\\
    &\quad\qquad-\frac{\binom{3}{1}2^{2\beta^2}2^{5\beta^2(n-1)}(2^{(20\beta^2-4)n}-1)}{(2^{2\beta^2}-2)^2(2^{6\beta^2}-4)(2^{20\beta^2-4}-1)}-\frac{2^{(11\beta^2-2)(n-1)}(2^{(14\beta^2-2)n}-1)}{(2^{2\beta^2}-2)(2^{14\beta^2-2}-1)}\Bigg]\nonumber\\
  &\quad\quad+2^{(25\beta^2-4)n}.
\end{align}
\endgroup

Thus, if $5\beta^2<1$ then
\begin{align}
  \frac{\mom_n(5,\beta)}{2^{5\beta^2n}}&\sim\frac{15\cdot2^{10 \beta^2+1}}{\left(2-2^{2 \beta^2}\right) \left(4-2^{6 \beta^2}\right)\left(16-2^{20 \beta^2}\right)}+\frac{15\cdot2^{20 \beta^2+2}}{\left(2-2^{2 \beta^2}\right)\left(4-2^{6 \beta^2}\right) \left(8-2^{12 \beta^2}\right) \left(16-2^{20\beta^2}\right)}\nonumber\\
  &\quad+\frac{15\cdot2^{16 \beta^2}}{\left(2-2^{2 \beta^2}\right)^2 \left(8-2^{12\beta^2}\right) \left(16-2^{20 \beta^2}\right)}
\end{align}
as $n\rightarrow\infty$.  Otherwise if $5\beta^2>1$ then, as $n\rightarrow\infty$,
\begingroup
\allowdisplaybreaks
\begin{align}
  &\frac{\mom_n(5,\beta)}{2^{(25\beta^2-4)n}}\nonumber\\
  &\quad\sim \frac{30 \left(2^{6\beta^2}-2\right) \left(2^{2 \beta^2}-1\right)}{\left(2^{2 \beta^2}-2\right) \left(2^{4 \beta^2}-2\right) \left(2^{6 \beta^2}-4\right) \left(2^{12 \beta^2}-2\right)}-\frac{15\cdot2^{4 \beta^2+1} \left(2^{2 \beta^2}-1\right)^2}{\left(2^{2 \beta^2}-2\right)^2 \left(2^{4 \beta^2}-2\right) \left(2^{16 \beta^2}-4\right)}\nonumber\\
  &\quad\quad+\frac{10 \left(2^{2 \beta^2}-1\right)}{\left(2^{2 \beta^2}-2\right) \left(2^{12 \beta^2}-2\right)}+\frac{15\cdot2^{6 \beta^2+1} \left(2^{2 \beta^2}-1\right)}{\left(2^{2 \beta^2}-2\right)^2 \left(2^{4 \beta^2}-2\right) \left(2^{18 \beta^2}-8\right)}\nonumber\\
  &\quad\quad+\frac{15\cdot2^{8 \beta^2+1} \left(2^{2 \beta^2}-1\right)}{\left(2^{2 \beta^2}-2\right)^2 \left(2^{6 \beta^2}-4\right) \left(2^{18 \beta^2}-8\right)}+\frac{60}{\left(2^{4 \beta^2}-2\right) \left(2^{6 \beta^2}-4\right) \left(2^{8 \beta^2}-2\right)}\nonumber\\
  &\quad\quad+\frac{20}{\left(2^{6 \beta^2}-2\right) \left(2^{8 \beta^2}-2\right)}+\frac{5}{2^{8 \beta^2}-2}-\frac{60 \left(2^{6 \beta^2}-2^{4 \beta^2}\right)}{\left(2^{2 \beta^2}-2\right) \left(2^{4 \beta^2}-2\right) \left(2^{8 \beta^2}-2\right) \left(2^{10 \beta^2}-4\right)}\nonumber\\
  &\quad\quad-\frac{15\cdot2^{2 \beta^2}}{\left(2^{2 \beta^2}-2\right) \left(2^{8 \beta^2}-2\right) \left(2^{10 \beta^2}-4\right)}-\frac{15\cdot2^{2 \beta^2+1}}{\left(2^{2 \beta^2}-2\right)^2 \left(2^{8 \beta^2}-2\right) \left(2^{10 \beta^2}-4\right)}\nonumber\\
  &\quad\quad+\frac{15\cdot2^{8 \beta^2+2}}{\left(2^{2 \beta^2}-2\right) \left(2^{6 \beta^2}-4\right) \left(2^{8 \beta^2}-2\right) \left(2^{12 \beta^2}-8\right)}+\frac{15\cdot2^{4 \beta^2}}{\left(2^{2 \beta^2}-2\right)^2 \left(2^{8 \beta^2}-2\right) \left(2^{12 \beta^2}-8\right)}\nonumber\\
  &\quad\quad-\frac{15\cdot2^{2 \beta^2+1} \left(2^{6 \beta^2}-2\right)}{\left(2^{2 \beta^2}-2\right) \left(2^{4 \beta^2}-2\right) \left(2^{6 \beta^2}-4\right) \left(2^{14 \beta^2}-4\right)}-\frac{5\cdot2^{2 \beta^2+1}}{\left(2^{2 \beta^2}-2\right) \left(2^{14 \beta^2}-4\right)}\nonumber\\
  &\quad\quad-\frac{15\cdot2^{6 \beta^2+2}}{\left(2^{4 \beta^2}-2\right) \left(2^{6 \beta^2}-4\right) \left(2^{14 \beta^2}-4\right)}-\frac{5\cdot2^{6 \beta^2+2}}{\left(2^{6 \beta^2}-2\right) \left(2^{14 \beta^2}-4\right)}-\frac{15\cdot2^{8\beta^2}}{\left(2^{8 \beta^2}-2\right) \left(2^{16 \beta^2}-4\right)}\nonumber\\
  &\quad\quad-\frac{15\cdot2^{8 \beta^2}}{\left(2^{2 \beta^2}-2\right) \left(2^{8 \beta^2}-2\right) \left(2^{16\beta^2}-4\right)}-\frac{15\cdot2^{8 \beta^2}}{\left(2^{2 \beta^2}-2\right)^2 \left(2^{8\beta^2}-2\right) \left(2^{16 \beta^2}-4\right)}\nonumber\\
  &\quad\quad+\frac{60 \left(2^{16\beta^2}-2^{14 \beta^2}\right)}{\left(2^{2 \beta^2}-2\right) \left(2^{4 \beta^2}-2\right) \left(2^{10\beta^2}-4\right) \left(2^{18 \beta^2}-8\right)}+\frac{15\cdot2^{12 \beta^2}}{\left(2^{2\beta^2}-2\right) \left(2^{10 \beta^2}-4\right) \left(2^{18 \beta^2}-8\right)}\nonumber\\
  &\quad\quad+\frac{15\cdot2^{12\beta^2+1}}{\left(2^{2 \beta^2}-2\right)^2 \left(2^{10 \beta^2}-4\right) \left(2^{18\beta^2}-8\right)}-\frac{15\cdot2^{10 \beta^2+1}}{\left(2^{2 \beta^2}-2\right)^2 \left(2^{6\beta^2}-4\right) \left(2^{20 \beta^2}-16\right)}\nonumber\\
  &\quad\quad-\frac{15\cdot2^{20 \beta^2+2}}{\left(2^{2\beta^2}-2\right) \left(2^{6 \beta^2}-4\right) \left(2^{12 \beta^2}-8\right) \left(2^{20\beta^2}-16\right)}-\frac{15\cdot2^{16 \beta^2}}{\left(2^{2 \beta^2}-2\right)^2 \left(2^{12\beta^2}-8\right) \left(2^{20 \beta^2}-16\right)}\nonumber\\
  &\quad\quad+\frac{15}{\left(2^{2 \beta^2}-2\right)\left(2^{8 \beta^2}-2\right)^2}+\frac{15}{\left(2^{2 \beta^2}-2\right)^2 \left(2^{8\beta^2}-2\right)^2}+\frac{15}{\left(2^{8 \beta^2}-2\right)^2}+1.
\end{align}
\endgroup

Observe that in all cases computed, in the limit $n\rightarrow\infty$, the leading order coefficient of $\mom_n(k,\beta)$ fails to be analytic in $\beta$.  Figures~\ref{fig:mom(2,b)}, \ref{fig:mom(3,b)}, \ref{fig:mom(4,b)}, and \ref{fig:mom(5,b)} plot the leading order coefficients of $\mom_n(k,\beta)$ for $k=2,3,4,5$ as $\beta$ varies.

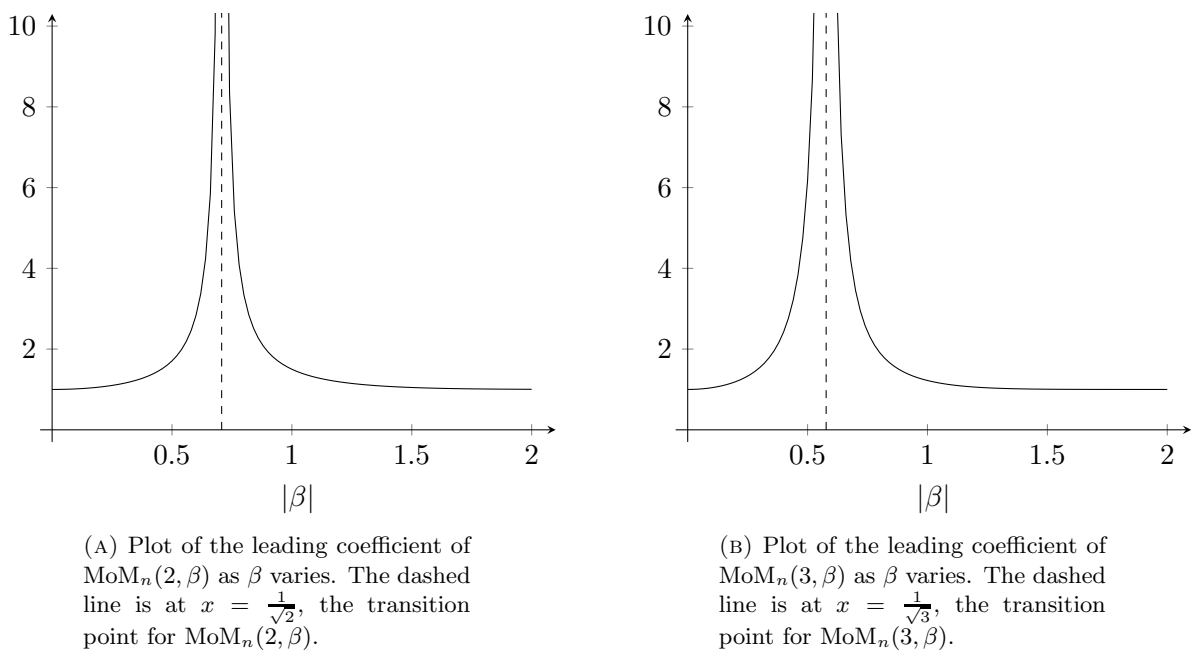
\begin{figure}
  \centering
  \subfloat[][Plot of the leading coefficient of $\mom_n(2,\beta)$ as $\beta$ varies.  The dashed line is at $x=\frac{1}{\sqrt{2}}$, the transition point for $\mom_n(2,\beta)$.]{
    \centering
    \begin{tikzpicture}[
        declare function={
          func(\x)= (\x < 0.7071) * (1/(2-2^(2*x*x))) + and(\x >= 0.7071, \x<= 0.7071) * (1/2) + (\x > 0.7071) * ((2^(2*x*x)-1)/(2^(2*x*x)-2));
        }
      ]
      \begin{axis}[
          axis x line=middle,
          axis y line=middle,
          ymin=-0.3,
          ymax=10.3,
          xmin=-0.05,
          xmax=2.1,
          domain=0:2,
          samples=101,
          xlabel near ticks,
          xlabel = $\abs{\beta}$
        ]
        \addplot[black] {func(x)};
        \addplot[black, dashed,domain=0:10.5] (0.7071,x);
      \end{axis}
    \end{tikzpicture}
    \label{fig:mom(2,b)}}
  \qquad
  \subfloat[][Plot of the leading coefficient of $\mom_n(3,\beta)$ as $\beta$ varies.  The dashed line is at $x=\frac{1}{\sqrt{3}}$, the transition point for $\mom_n(3,\beta)$.]{
    \centering
    \begin{tikzpicture}[
        declare function={
          func(\x)= (\x < 0.5774) * (3*(2^(2*x*x))/((2^(6*x*x)-4)*(2^(2*x*x)-2))) + and(\x >= 0.5774, \x<= 0.5774) * (0) + (\x > 0.5774) * (1+(3*(2^(6*x*x)-2))/((2^(4*x*x)-2)*(2^(6*x*x)-4)));
        }
      ]
      \begin{axis}[
          axis x line=middle,
          axis y line=middle,
          ymin=-0.3,
          ymax=10.3,
          xmin=-0.05,
          xmax=2.1,
          domain=0:2,
          samples=101,
          xlabel near ticks,
          xlabel = $\abs{\beta}$
        ]
        \addplot[black] {func(x)};
        \addplot[black, dashed,domain=0:10.5] (0.5774,x);
      \end{axis}
    \end{tikzpicture}
    \label{fig:mom(3,b)}}
  \caption[]{Figures showing the leading coefficients of $\mom_n(2,\beta)$ and $\mom_n(3,\beta)$ as $\beta$ varies.}
\end{figure}

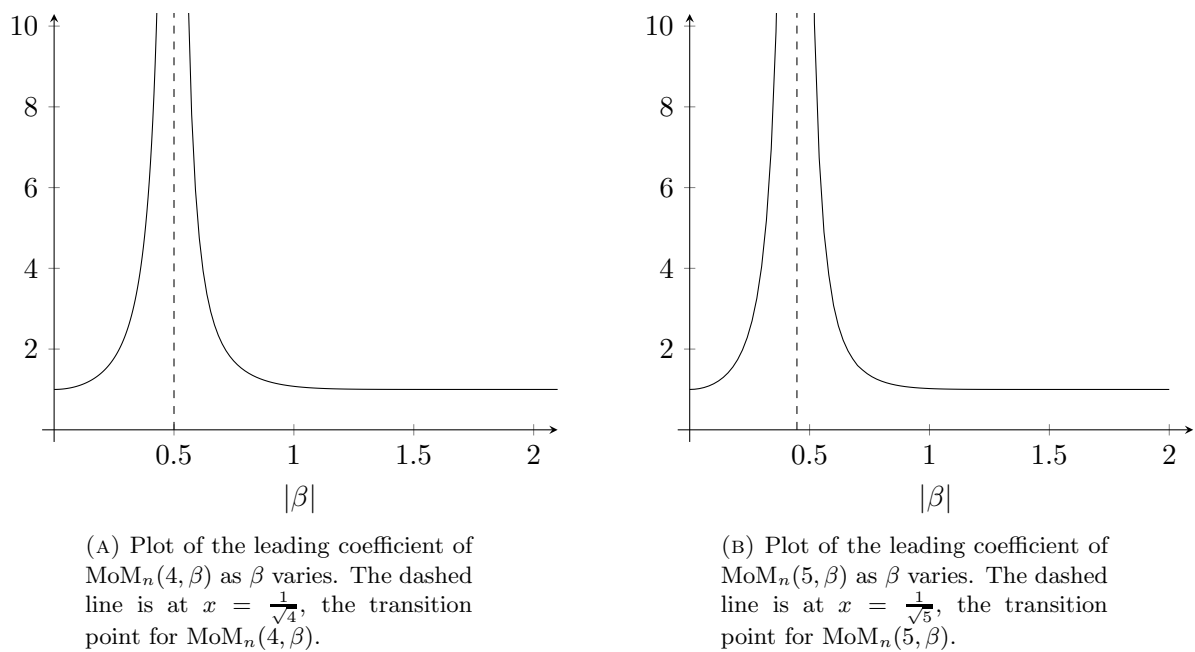
\begin{figure}
  \centering
  \subfloat[][Plot of the leading coefficient of $\mom_n(4,\beta)$ as $\beta$ varies.  The dashed line is at $x=\frac{1}{\sqrt{4}}$, the transition point for $\mom_n(4,\beta)$.]{
    \centering
    \begin{tikzpicture}[
        declare function={
          func(\x)= (\x < 0.501) * ((12*2^(8*x*x))/((2^2-2^(6*x*x))*(2-2^(2*x*x))*(2^3-2^(12*x*x)))+(6*2^(8*x*x-4))/((2-2^(2*x*x))*(2-2^(2*x*x))*(2^(4*x*x)-2^(16*x*x-3)))) + and(\x >= 0.499, \x<= 0.501) * (0) + (\x > 0.499) * ((2^(28*x*x)-80+2^(36*x*x)+2^(3+(6*x*x))-2^(5+(8*x*x))+(17*(2^(2+(10*x*x))))-(5*(2^(2+(12*x*x))))+2^(5+(14*x*x))-2^(1+(16*x*x))-(7*(2^(1+(18*x*x))))+2^(4+(20*x*x))-(5*(2^(1+(22*x*x))))-2^(3+(24*x*x))-2^(2+(26*x*x))+2^(1+(30*x*x)))/((2^(6*x*x)-2)*(2^(8*x*x)-2)*(2^(10*x*x)-4)*(2^(12*x*x)-8)));
        }
      ]
      \begin{axis}[
          axis x line=middle,
          axis y line=middle,
          ymin=-0.3,
          ymax=10.3,
          xmin=-0.05,
          xmax=2.1,
          domain=0:2,
          samples=101,
          xlabel near ticks,
          xlabel = $\abs{\beta}$,
          line join=round
        ]
        \addplot[black,domain=0:0.49] {func(x)};
        \addplot[black,domain=0.51:2.1] {func(x)};
        \addplot[black, dashed,domain=0:10.5] (0.5,x);
      \end{axis}
    \end{tikzpicture}
    \label{fig:mom(4,b)}}
  \qquad
  \subfloat[][Plot of the leading coefficient of $\mom_n(5,\beta)$ as $\beta$ varies.  The dashed line is at $x=\frac{1}{\sqrt{5}}$, the transition point for $\mom_n(5,\beta)$.]{
    \centering
    \begin{tikzpicture}[
        declare function={
          func(\x)= (\x < 0.449) * (((30*(2^(16*x*x)))/(2*(8-2^((12*x*x)))*(2-2^(2*x*x))*(2-2^(2*x*x))*(16-2^(20*x*x))))+((60*(2^(20*x*x)))/((4-2^(6*x*x))*(2-2^(2*x*x))*(8-2^(12*x*x))*(16-2^(20*x*x))))+((30*(2^(10*x*x)))/((4-2^(6*x*x))*(2-2^(2*x*x))*(16-2^(20*x*x))))) + and(\x >= 0.447, \x<= 0.448) * (0) + (\x > 0.447) * (5/(2^(8*x*x)-2) + 20/((2^(6*x*x)-2)*(2^(8*x*x)-2)) - 20*(2^(6*x*x))/((2^(6*x*x)-2)*(2^(14*x*x)-4))+60/((2^(6*x*x)-4)*(2^(4*x*x)-2)*(2^(8*x*x)-2))-(60*(2^(6*x*x)))/((2^(6*x*x)-4)*(2^(4*x*x)-2)*(2^(14*x*x)-4)) - (60*(2^(6*x*x)-2^(4*x*x)))/((2^(10*x*x)-4)*(2^(4*x*x)-2)*(2^(2*x*x)-2)*(2^(8*x*x)-2)) + (60*(2^(16*x*x)-2^(14*x*x)))/((2^(10*x*x)-4)*(2^(4*x*x)-2)*(2^(2*x*x)-2)*(2^(18*x*x)-8))+(60*2^(8*x*x))/((2^(2*x*x)-2)*(2^(6*x*x)-4)*(2^(12*x*x)-8)*(2^(8*x*x)-2)) - (60*2^(20*x*x))/((2^(2*x*x)-2)*(2^(6*x*x)-4)*(2^(12*x*x)-8)*(2^(20*x*x)-16))+(15*2^(4*x*x))/((2^(2*x*x)-2)*(2^(2*x*x)-2)*(2^(12*x*x)-8)*(2^(8*x*x)-2))-(15*(2^(16*x*x)))/((2^(2*x*x)-2)*(2^(2*x*x)-2)*(2^(12*x*x)-8)*(2^(20*x*x)-16))+15/((2^(2*x*x)-2)*(2^(2*x*x)-2)*(2^(8*x*x)-2)*(2^(8*x*x)-2))-(15*(2^(8*x*x)))/((2^(2*x*x)-2)*(2^(2*x*x)-2)*(2^(8*x*x)-2)*(2^(16*x*x)-4))+15/((2^(8*x*x)-2)*(2^(8*x*x)-2))-(15*(2^(8*x*x)))/((2^(8*x*x)-2)*(2^(16*x*x)-4))+15/((2^(2*x*x)-2)*(2^(8*x*x)-2)*(2^(8*x*x)-2))-(15*(2^(8*x*x)))/((2^(2*x*x)-2)*(2^(8*x*x)-2)*(2^(16*x*x)-4))-(30*(2^(2*x*x)))/((2^(2*x*x)-2)*(2^(2*x*x)-2)*(2^(10*x*x)-4)*(2^(8*x*x)-2))+(30*(2^(12*x*x)))/((2^(2*x*x)-2)*(2^(2*x*x)-2)*(2^(10*x*x)-4)*(2^(18*x*x)-8))-(15*(2^(2*x*x)))/((2^(2*x*x)-2)*(2^(10*x*x)-4)*(2^(8*x*x)-2))+(15*(2^(12*x*x)))/((2^(2*x*x)-2)*(2^(10*x*x)-4)*(2^(18*x*x)-8))+(30*(2^(2*x*x)-1)*(2^(6*x*x)-2))/((2^(2*x*x)-2)*(2^(6*x*x)-4)*(2^(4*x*x)-2)*(2^(12*x*x)-2))-(30*(2^(2*x*x)-1)*(2^(2*x*x)-1)*2^(4*x*x))/((2^(2*x*x)-2)*(2^(2*x*x)-2)*(2^(4*x*x)-2)*(2^(16*x*x)-4))+(30*(2^(2*x*x)-1)*2^(8*x*x))/((2^(2*x*x)-2)*(2^(2*x*x)-2)*(2^(6*x*x)-4)*(2^(18*x*x)-8))+(10*(2^(2*x*x)-1))/((2^(2*x*x)-2)*(2^(12*x*x)-2))-(30*(2^(6*x*x)-2)*2^(2*x*x))/((2^(2*x*x)-2)*(2^(6*x*x)-4)*(2^(4*x*x)-2)*(2^(14*x*x)-4))+(30*(2^(2*x*x)-1)*2^(6*x*x))/((2^(2*x*x)-2)*(2^(2*x*x)-2)*(2^(4*x*x)-2)*(2^(18*x*x)-8))-(30*2^(10*x*x))/((2^(2*x*x)-2)*(2^(2*x*x)-2)*(2^(6*x*x)-4)*(2^(20*x*x)-16))-(10*2^(2*x*x))/((2^(2*x*x)-2)*(2^(14*x*x)-4))+1);
        }
      ]
      \begin{axis}[
          axis x line=middle,
          axis y line=middle,
          ymin=-0.3,
          ymax=10.3,
          xmin=-0.05,
          xmax=2.1,
          domain=0:2,
          samples=101,
          xlabel near ticks,
          xlabel = $\abs{\beta}$,
          line join=round
        ]
        \addplot[black] {func(x)};
        \addplot[black, dashed,domain=0:10.5] (0.44721,x);
      \end{axis}
    \end{tikzpicture}
    \label{fig:mom(5,b)}}
  \caption[]{Figures showing the leading coefficients of $\mom_n(4,\beta)$ and $\mom_n(5,\beta)$ as $\beta$ varies.}
\end{figure}

\end{document}